\documentclass{llncs}
\usepackage[german,english]{babel}
\usepackage{times,theorem}
\usepackage{latexsym,color,graphics}
\usepackage{diagrams}
\usepackage{url}
\usepackage{epsfig}
\usepackage{amssymb}
\usepackage{amsmath}
\usepackage{amsfonts}

\begingroup
\catcode`\~=11
\gdef\urltilde{\lower 0.6ex\hbox{~}}
\endgroup

 \newcommand{\B}{\mathcal{B}}
\newcommand{\C}{\mathcal{C}} 
\newcommand{\E}{\mathcal{E}} \newcommand{\F}{\mathcal{F}}
 
\newcommand{\I}{\mathcal{I}} 
\newcommand{\K}{\mathcal{K}} \renewcommand{\L}{\mathcal{L}}
\newcommand{\M}{\mathcal{M}} \newcommand{\N}{\mathcal{N}}
 \renewcommand{\P}{\mathcal{P}}
 \newcommand{\R}{\mathcal{R}}
 \newcommand{\T}{\mathcal{T}}
 
\newcommand{\W}{\mathcal{W}}

\title{Reduction of Many-valued  into Two-valued Modal Logics}
\author{Zoran Majki\'c}
\author{Zoran Majki\'c}
\institute{ International Society for Research in Science and Technology\\
 PO Box 2464 Tallahassee, FL 32316 - 2464 USA\\
 \email{majk.1234@yahoo.com},\\
~~~~http://zoranmajkic.webs.com/}
\authorrunning{Zoran Majki\'c}

\newtheorem{propo}{Proposition}

\newcount\pdfoutput
\begin{document}


\maketitle

\begin{abstract}
In this paper we develop a 2-valued reduction of many-valued logics,
 into  2-valued multi-modal logics. Such an approach is
based on the contextualization of many-valued logics with the
introduction of higher-order Herbrand interpretation types, where we
explicitly introduce the coexistence of a set of algebraic truth
values of original many-valued logic, transformed as parameters (or
possible worlds), and the set of classic two logic values. This
approach is close to the approach used in annotated logics, but
offers the possibility of using  the standard semantics based on
Herbrand interpretations. Moreover, it uses the properties of the
higher-order Herbrand types, as their fundamental nature is based on
autoreferential Kripke semantics where the possible worlds are
algebraic truth-values of original many-valued logic. This
autoreferential Kripke  semantics, which has the possibility of
flattening higher-order Herbrand interpretations into ordinary
2-valued Herbrand interpretations, gives us a  clearer  insight into
the relationship between many-valued and 2-valued multi-modal
logics. This methodology is applied to the class of many-valued
Logic Programs, where reduction is done in a structural way, based
on the logic structure (logic connectives) of original many-valued
logics. Following this, we generalize the reduction to general
structural many-valued logics, in an abstract way, based on Suszko's
informal non-constructive idea. In all cases, by using developed
2-valued reductions we obtain a kind of non truth-valued modal
meta-logics, where two-valued formulae are modal sentences obtained
by application of particular modal operators to original many-valued
formulae.
\end{abstract}

\section{Introduction}
 A significant number of real-world applications in Artificial Intelligence
 have to deal with partial, imprecise and uncertain information, and that
 is the principal reason for introducing the non-classic many-valued
 logics,  for example, fuzzy,  bilattice-based and
 paraconsistent logics, etc..\\
 In such cases we associate some \emph{degree of belief} to ground atoms,
which can be simple probability, probability interval, or other more
complex data structures, as for example in Bayesian logic programs
where for  different kinds of atoms we associate also different
(that is, different from probability) kinds of \emph{measures}. The
many-valued logics with a set of such measures (that is, 'algebraic
truth-values') are one of the
main tools that we can use for such  applications. \\
 The reduction of many-valued logics into the standard 2-valued logic was
considered by Suszko
 \cite{Susz75}, where he illustrated how Lukasiewicz's 3-valued
 logic could be given a 2-valued, non truth-functional, semantics.
 The main point, according to Suszko, is to make a distinction between
 the \emph{algebraic truth-values} in $\W$ of many-valued logics, which were
 supposed to play a merely referential role, while only two \emph{logical
 truth-values} in
$\textbf{2} = \{0,1\}$  ($0$ for false and $1$ for true value) would
really exist. It is also based on the fact that the abstract logic
is based on a \emph{consequence relation} that is bivalent: given a
set of logic formulae $S$, a formula $\phi$
 can be inferred  from $S$ or not, that is, the answer to the question
 "if $\phi$ is inferred from $S$"
 can  only be 'Yes' or 'No'.\\
 This point of view for 'logic values' is also
 considered correct  by other authors, and it is also applied  in the case of an ontological encapsulation \cite{Majk04on}
 of many-valued algebraic logic programs into 2-valued logic
 programs. Moreover, in a 2-valued reduction,
 for any propositional formula $\phi$ that has an 'algebraic truth-value' $\alpha$,
we can consider a 2-valued meta-sentence  "the truth-value of $\phi$
is $\alpha$",
 i.e., $t(\phi,\alpha)$ where $t$ is a binary predicate for true sentences and $\alpha \in \W$ an algebraic truth value.
  In order to avoid a second order logic with the formula $t(\phi,\alpha)$, we
   can transform it into a First Order (FO) formula $[\alpha] \phi$ instead, with the introduction of
   a modal connective  $[\alpha]$ as in \cite{Majk09BS}.\\
 Suszko's thesis for the reduction of every tarskian (monotonic)
n-valued logic into a 2-valued logic is based on this division of a
set of logic values into a subset of designated and undesignated
elements, but it is quite a non-constructive result. In fact, he
does not explain  how he obtained a 2-valued
semantics, or how such a procedure could be effectively applied.\\
In the paper by D.Batens \cite{Bate82}, the author proposes  a sort
of binary print of the algebraic truth-values for the 2-valued
reduction, where each truth-value is to be put into one-to-one
correspondence with one element of a set of conveniently long
'equivalent' sequences of 0's and 1's. This method is similar to
what had been proposed by D.Scott a decade before \cite{Scot74}. But
this method is not universally applicable and thus can not be
effectively used. Some other authors argued against Suszko's thesis
\cite{CoBB96} using examples of paraconsistent logic and
Malinowski's inferential many-valuedness. But recently in
\cite{CCCM05}, based on Suszko's observations on complementarity of
designated and undesignated elements,  a method was exhibited  for
the effective implementation of Suszko's reduction to a subclass of
finite-valued truth-functional logics, whose truth-values satisfy
the particular assumption of separability, where the 'algebraic
truth-values' can be individualized by means of the linguistic
resources of the logic. What is important for the present work is
that they show that a reduction of truth-functional many-valued
logic into 2-valued logic will simply make it lose truth-functionality: in fact, our  transformation will result in \emph{modal} logics.\\
Consequently, the main contribution of this paper is to use \emph{a
constructive} approach to Suszko's method, and to exhibit a  method
for the effective implementation of 2-valued reduction \emph{for all
kinds} of many-valued logics. It avoids the necessity of dividing
(in problematic way based on subjective opinions) a set of algebraic
truth-values into designated and undesignated disjoint subsets in
order to define the satisfaction relation (i.e., entailment),  by
using the valuations (model-theoretic semantics): the entailment $S
\models \phi$ means that every model (valuation) of $S$ is a model
of $\phi$. For example, any rule in a many-valued logic program $A
\leftarrow B_1,..., B_n$ is \emph{satisfied} if, for a given
valuation $v$, the algebraic truth-value of the head is greater than
the value of the body, i.e., if $v(A) \geq v(B_1 \wedge...\wedge
B_n)$. More discussion about this approach  can be found in a new
representation theorem for many-valued logics
\cite{Majk06th}. \\
 Consequently, in what follows we will consider a  possible embedding of  these
 many-valued logics into  2-valued logic, in order to understand a basic connection between them and the well investigated families of
 2-valued sublanguages (logics) of the first order logic language. In the past, some approaches were  made in this
 direction, as ad-hoc logics (for example, annotated logic), but
   without the real purpose of investigating  this issue. We will consider the following two approaches for \emph{predicate} many-valued logics
   (the propositional version can be considered as a special case, when all predicate symbols have a zero
   arity): the first one introduces \emph{unary} modal operator for
   each truth value of original many-valued logic; the second
   approach introduces the \emph{binary} modal operator for each
   binary truth-valued logic operator (conjunction, disjunction,
   implication) of original many-valued logic.
   Both of them will transform an original truth-functional many-valued logic
   into \emph{non truth-functional} 2-valued modal logic, as
   follows:
   \\
 1. In ~\cite{KiSu91} it is shown that Fitting's 3-valued bilattice logic can be
 embedded into an Annotated Logic Programming that is
 computationally very complex and has a non standard (that is, Herbrand based) interpretation. In what follows we will use the
 syntactic annotation for  many-valued logic programs, with a set of logic values in $\W$, where a rule  of the form
 $~A:f(\beta_1,..,\beta_n) \leftarrow B_1:\beta_1
 ,...,B_n:\beta_n~$,  asserts "the 'truth' of the atom $A$
 is at least (or is in) $f(\beta_1,..,\beta_n) = \beta_1 \wedge ...\wedge \beta_n$ (the result of the many-valued logic conjunction of logic
 values $\beta_i \in \W$).\\
  We will extend this consideration by introducing a \emph{contextual}
 logic, which is a syntax variation of the annotated logic, where
 instead of annotated atoms $B:\beta$ we will use a couple $(B, \beta)$ that is a more practical set-based denotation
 and can have the Herbrand interpretations. It is the
 fundamental and  first step when we try to transform a
 many-valued logic into positive 2-valued logic programs with classical conjunction and implication, where we will use
 modal atoms $[\beta]B$, ($[\beta]$ denotes a universal modal operator), instead of annotated atoms.
 As we will see, such a contextualization of  many-valued logic
 programs generates the higher-order Herbrand interpretations.\\
 2. The ontological embedding
 \cite{MajkC04}
 into the syntax of new encapsulated many-valued logic
 (in some sense meta-logic for a many-valued bilattice logic)
 is a 2-valued, and can be seen as a flattening of a many-valued logic, where an algebraic truth-value
  $\beta \in \W$ of an original ground atom $r(c_1,..,c_k)$ is deposited into the logic attribute of a new predicate $r_F$,
  obtained by an extension of the old predicate $r$, so that we obtain the 'flattened' 2-valued
  ground atom $r_F(c_1,..,c_k, \beta)$. In that case, we will obtain the positive multi-modal logic programs
  with binary modal operators for conjunction, disjunction and implication and unary modal operator for negation.\\
These two \emph{knowledge invariant} 2-valued logic transformations
of the original many-valued logic program are mutually inverse: we
can consider the annotations as the contexts for the original atoms
of the logic theory. Such a context sensitive application, with
higher-order Herbrand models, can be transformed (that is,
\emph{flattened}) into the logic theories with basic (ordinary)
Herbrand interpretations, by enlarging the original predicates with
new attributes that characterize the properties of the context: in
this way a context also becomes  a part of the language of a logic
theory, that is, it becomes visible.\\
 The inverse of a flattening is a predicate compression
 \cite{Majk06FM}.
In this paper we will implicitly consider only a compression of the
logic attribute of the flattened predicates obtained during
ontological encapsulation of a many-valued logic program: the
obtained compressed predicates are identical to the predicates from
the original many-valued logic program, but the value for their
ground atoms is not a value of a basic set of algebraic truth-values
in $\W$ but a \emph{function} (higher-order value type) in
$\textbf{2}^{\W}$ (the set of all functions from $\W$ to
$\textbf{2}$).
 A contextualization of a
many-valued logic is equivalent to the compression of logical
variables of the flattened versions of many-valued logic programs.

Both approaches above are different from somewhat similar procedures
investigated by Pavelka in \cite{Pave79} by expansion of
propositional Lukasiewicz's logic with a truth-constant
$\overline{\beta}$ for every real value $\beta \in [0,1]$, and
successively refined by H$\acute{\textrm{a}}$jek in \cite{Haje98}
and brought to first order predicate systems in
\cite{Nova90,EGNo07}. In fact in the first approach above we
introduce not logic constants (\emph{nullary} logic operators), but
\emph{unary} modal operators for every truth-value, while in the
second approach above we introduce only new k-ary ($k \geq 1$)
built-in functions obtained from a semantic reflection of
many-valued Herbrand interpretations of predicate many-valued logics
and we enlarge the domain of values of the original logic by the set
of algebraic truth-values in $\W$.
\\The main \emph{motivation} of this work is a theoretical investigation
of the possibility of reducing a many-valued into a standard
2-valued logic. It is not our aim to replace the original
many-valued logics, which are more intuitive and natural
representations used in practice. But we would like to obtain the
2-valued reductions as a canonical form for the whole family of
various many-valued logics, where we can investigate their common
properties and make  comparisons between them. So, the main
\emph{contribution} of this article is that we present this possible
canonical reduction of any many-valued into 2-valued multi-modal
logic, and the possibility of reusing the rich quantity  of results
discovered for modal logics. In this way we also define  the
upper limit of the expressive power for any possible many-valued logic. \\
\textbf{Remark:} In what follows we are interested in general
many-valued algebras, based on a lattice $(\W, \leq, \wedge, \vee)$
of truth values (where ordering $\leq$ is interpreted as truth
ordering of logic values), where the meet $\wedge$ and join $\vee$
operators are the algebraic counterparts of logic conjunction and
disjunction respectively, and extended by other unary operators (for
example, by many-valued logic negation) and binary operators (for
example, by many-valued logic implication). We will denote by $0$
and $1$ the bottom and top elements respectively of such a lattice
$\W$ (if $\W$ is not a bounded lattice then we will add to it these
two elements).
Thus we are able to reduce a bounded lattice of a many-valued logic
$\W$ into the classic 2-valued logic with the set of logic values in
$\textbf{2} = \{0, 1\} \subset \W$ (where $\textbf{2}$ is a complete
sublattice of $\W$), in the way that the many-valued operators
defined in a bounded lattice $\W$  are reduced, by this two-valued
reduction, into the classical 2-valued logic operators (the
conjunction, disjunction, negation and material implication).
Because of that, the only restriction for many-valued negation
operator $\sim$ is that $\sim 0 = 1$ and $\sim 1 = 0$, such that it
is antitonic (i.e., satisfies De Morgan laws between the conjunction
and disjunction).
 The set of many-valued logic
connectives will be denoted by $\Sigma$. Two unrelated elements $a,
b \in \W$ will be denoted by $a \bowtie b$. In order to avoid
confusion between many-valued and 2-valued conjunction and
disjunctions,
where necessary, for 2-valued connectives we will use $\bigwedge$ and  $\bigvee$ symbols respectively.\\
  This paper follows the following plan: \\After a short
  introduction for 2-valued multi-modal logics, in Section 2 we present a theory for higher-order Herbrand interpretation types (and its
 correspondent flattening into the ordinary Herbrand
 interpretations) obtained in a process of contextualization by relativizing the truth
 (and falsity) of a logic formulae to a given context (or "possible
 world"). We show that this is a pre-modal development for logics and
 can be used directly  to define  2-valued concepts with
 Kripke semantics.
 In Section 3 we present a number of significative examples for
 many-valued logics, and show how they can be contextualized in
 order to be able to introduce the logic values of a many-valued
 logics as particular 'logic objects' into the  language of this contextual logic.
 The result of this contextualization (which renders visible logic
 values of a many-valued logic) is that the atoms in a Herbrand base
 have the higher-order logic values: a contextual logic has the higher-order Herbrand
 interpretations.
  We show how these higher-order Herbrand model types can be equivalently considered
 as multi-modal  Kripke models, where a set of possible worlds is taken from the structure of these higher-order types.
  In Section 4 we show how these techniques can be applied to
 many-valued Logic Programs, and we show that they can be
 equivalently transformed into 2-valued multi-modal Logic Programs.
 We consider two kinds of transformations: the first one by introducing the set of unary modal operators for each algebraic logic value,
  and the second one by introducing binary modal operators in the place of the original binary many-valued logic operators.\\
 Finally, in Section 5 we develop an abstract  method for a
 2-valued reduction of (general) many-valued logics, transforming
  Suszko's non-constructive idea into a formal method. This
 reduction results in a non truth-functional 2-valued modal
 meta-logic, where 2-valued sentences are obtained by applying
 specific modal operators to original many-valued logic formulae.
 %
%
\subsection{Introduction to predicate multi-modal logic}
%

A predicate multi-modal logic, for a language with a set of
predicate symbols $r \in P$ with arity $ar(r)\geq 0$ and a set of
functional symbols $f \in F$ with arity $ar(f) \geq 0$, is a
standard predicate logic extended by a \emph{finite} number of
universal modal operators $\Box_i, i \geq 1$. In this case we do not
require that these universal modal operators are normal modal (that
is, monotonic and multiplicative) operators as in a standard setting
for  modal logics, but we do require that they have the same
standard Kripke semantics. In a standard Kripke semantics each modal
operator $\Box_i$ is defined by an accessibility binary relation
$\R_i \subseteq \W \times \W$ in a given set of possible worlds
$\W$. A more exhaustive and formal introduction to modal logics and
their Kripke models can  easily be found in the literature, for
example in \cite{BBWo06}. Here  only a short version will be given,
in order to clarify the definitions used in the next paragraphs.\\
In what follows we  denote by $A\Rightarrow B$, or $B^A$, the set of
all functions from $A$ to $B$, and by $A^n$ a n-folded cartesian
product $A \times ...\times A$ for $n \geq 1$.\\
We define the set of terms of this predicate modal logic  as
follows:
 all variables $x \in Var$, and constants $d \in S$ are terms; if $f \in F$ is a functional symbol of arity $k = ar(f)$ and $t_1,..,
t_k$ are terms, then $~f(t_1,..,t_k)$ is a term. We denote by $\T_0$ the set of all ground (without variables) terms. \\
An atomic formula (atom) for a predicate symbol $r \in P$ with arity
$k = ar(r)$ is an expression $r(t_1,...,t_k)$, where $t_i, i
=1,...,k$ are terms. Herbrand base $H$ is a set of all ground atoms
(atoms without variables). More complex formulae, for a predicate
multi-modal logic, are obtained as a free algebra obtained from the
set of all atoms and usual set of classic 2-valued binary logic
connectives in $ \{\wedge, \vee, \rightarrow\}$ for conjunction,
disjunction and implication respectively (negation of a formula
$\phi$, denoted by $\neg \phi$ is expressed by $\phi \rightarrow 0$,
where $0$ is used for an inconsistent formula (has constantly value
$0$ for every valuation)), and  a number of unary universal modal
operators $\Box_i$. We define $\N = \{1,2,..., n\}$ where $n$ is a
maximal arity of symbols in the finite set $P \bigcup F$.
 \begin{definition} \label{def:KripSem}
 We denote by $\M = (\W, \{$$
{\R}_i~ | ~1 \leq i \leq k \}, S, V)$ a multi-modal  Kripke model
with finite $k \geq 1$ modal operators with
 a set of possible worlds $\W$, the accessibility relations ${\R}_i \subseteq \W  \times \W$,
 non empty set of individuals $S$, and    a function $~~V:\W\times (P \bigcup F) \rightarrow {\bigcup}_{n \in \N}
(\textbf{2}\bigcup S)^{S^n}$, such that for any world $w \in \W$,\\
1. For any functional letter $f \in F$, $~V(w,f):S^{ar(f)}
\rightarrow S~$ is a
function (interpretation of $f$ in $w$).\\
2. For any predicate letter $r \in P$, the function
$~V(w,r):S^{ar(r)} \rightarrow \textbf{2}~$ defines the extension of
$r$ in a world $w$, $~~\|r\| = \{ \textbf{d}  = <d_1,...,d_k> \in
S^k~|~ k = ar(r), V(w,r)(\textbf{d}) = 1 \}$.
 \end{definition}
 For any formula $\varphi$  we define  $~~{\M} \models_{w,g}~\varphi~$ iff $~\varphi$ is satisfied in a world $w \in \W$ for
 a given assignment $g:Var \rightarrow S$. For example, a given atom $r(x_1,...,x_k)$ is satisfied in $w$ by assignment $g$,
 i.e., $~{\M} \models_{w,g}~r(x_1,...,x_k),~$ iff $~V(w,r)(g(x_1),...,g(x_k)) = 1$.\\
 The Kripke semantics is extended to all formulae as follows:
 \\$~~{\M} \models_{w,g}~ \varphi \wedge \phi~~~$ iff $~~~{\M} \models_{w,g}~ \varphi~$ and $~{\M} \models_{w,g}~
 \phi~$,
 \\$~~{\M} \models_{w,g}~ \varphi \vee \phi~~~$ iff $~~~{\M} \models_{w,g}~ \varphi~$ or $~{\M} \models_{w,g}~
 \phi~$,
 \\$~~{\M} \models_{w,g}~ \varphi \rightarrow \phi~~~$ iff $~~~{\M} \models_{w,g}~ \varphi~$ implies $~{\M} \models_{w,g}~
 \phi~$,
\\ $~~{\M} \models_{w,g}~\Box_i \varphi~~~$ iff $~~~\forall w'((w,w')
\in {\R}_i $ implies ${\M} \models_{w',g}~ \varphi~)~$.\\ The
existential
modal operator $\diamondsuit_i $ is equal to $\neg \Box_i \neg$.\\
A formula $\varphi$ is said to be \emph{true in a model} ${\M}$ if
 for each assignment function $g$
and possible world $w$, ${\M} \models_{w,g}~\varphi$. A formula is
said to be \emph{valid} if it
is true in each model.\\
We denote by $|\phi/g| = \{w~|~\M  \models_{w,g'}~\phi/g\}$ the set
of all worlds where the ground formula $\phi/g$ (obtained from
$\phi$ and an assignment $g$) is satisfied.
%
\section{Contextualization: Higher-order Herbrand interpretation types \label{Section:HiHerbrand}}
The higher-order types of Herbrand interpretations for many-valued
logic programs, where we are not able to associate a fixed logic
value to a given ground atom of a Herbrand base  but  a function in
a given functional space, often arise  in practice when we have to
deal with uncertain information. In such cases we associate some
\emph{degree of belief} to ground atoms, which can be simple
probability, probability interval, or other more complex data
structures, as for example in Bayesian logic programs where for a
different kind of atoms we may associate different
 kinds of \emph{measures} as well.\\
But we can see  approximate (uncertain) information as a kind of
\emph{relativization} of truth values for sentences as follows. Let
$H$ be a Herbrand base for a logic program that handles the
uncertain information, and $r(\textbf{d})$ a ground atom in $H$ that
logically defines a particular fact for which we have only an
approximated information about when it happened. Thus, this atom
$r(\textbf{d})$ is no longer \emph{absolutely} true or false, but
rather its truth depends on the approximate temporal information
about this fact: in some time points it can be true, in other it can
be false. If we consider such a temporal approximation as a
\emph{context} for this ground fact $r(\textbf{d}) \in H$, then we
obtain that the truth of $r(\textbf{d})$ is a function from the time
to the ordinary set of truth values $\textbf{2} = \{0,1\}$.
Consequently, the truth values of ground atoms in this Herbrand base
are the functions, that is, they have a \emph{higher-order type}
(this term is taken from the typed lambda calculus) with respect to
the set $\textbf{2}$ of truth constants. Intuitively, the
approximated information is relativized to its context, and such a
context further specifies the
semantics for this uncertain information.\\
The \emph{contextualization} is a kind of \emph{pre-modal} Kripke
modeling: in fact, if we consider a context as a Kripke "possible
world", then the relativization of the truth to particular contexts
is equivalent to Kripke semantics for a modal logic where the truth
(or falsity) of the formulae is relativized to possible worlds. In
fact, as we will see in what follows, the higher-order Herbrand
models obtained by contextualization are precursors for an
introduction of 2-valued epistemic concepts, that is, for a
development of
(absolute) 2-valued logics, and it explains their role in a 2-valued reduction of many-valued logics. \\
The higher-order Herbrand interpretations of logic programs produce
the models where the true values for ground atoms are not truth
constants but functions:
\begin{definition} \label{def:HOHerbrand}  \cite{Majk06} \textsc{Higher-order Herbrand interpretation types}: \\Let $H$ be a Herbrand
base, then, the higher-order Herbrand interpretations are defined by
$~I:H\rightarrow T~$, where $T $ is a functional space
$W_1\Rightarrow (...(W_n \Rightarrow \textbf{2})...)$, denoted also
as $(...((\textbf{2}^{W_n})^{W_{n-1}})...)^{W_1}$, and $W_i, ~ i \in
[1,n],~n\geq 1 $ are the sets of parameters (the values of given
domains).
 In the case  $n = 1, \W = W_1$,  $T = (\W \Rightarrow
\textbf{2})$, we will  denote this interpretation by
$~I:H\rightarrow \textbf{2}^{\W}~$.
\end{definition}
In \cite{Majk07hc} there has been developed a general method of
constructing  2-valued autoepistemic language concepts  for each
many-valued ground atom with higher-order Herbrand interpretation
given in Definition \ref{def:HOHerbrand}, for which we would like
 to have a correspondent 2-valued logic language concept. The number of such atomic concepts to be used in
 the applications is always a \emph{finite} subset $H_M$ of $M$ elements of the Herbrand
 base $H$.
\begin{definition} \cite{Majk07hc} \textsc{Epistemic Concepts}:
\label{def:concept} Let $\overline{H_M}$ be a finite sequence of N
ground atoms in $H$, $H_M$ a set of elements in $\overline{H_M}$,
and $~i_N:H_M \hookrightarrow H$ be an inclusion
 mapping for this finite subset of ground atoms.  We define the bijection $~i_C:H_M \simeq C_M$,
 with the set of derived concepts $C_M = \{ \Box_i A | A = \pi_i(\overline{H_M}), 1 \leq i \leq M \}$, where $\pi_i$ is i-th projection,
 such that for any ground atom $A = \pi_i(\overline{H_M})$, $~~i_C(A) =
\Box_i A$.
\end{definition}
The idea of how to pass to the possible-world Kripke semantics for
modal operators $\Box_i$, used above for an epistemic definition of
concepts, is
as follows: we define the set\\
   $~~~Q_{i} = \{\textbf{w}~|~~r(\textbf{d})
 = \pi_i(\overline{H_M}) \in H$ and $~I(r(\textbf{d}))(\textbf{w})
  =1\}$.\\ It is easy to verify that $Q_{i}$ is the set of all points $\textbf{w} \in \W$ where the ground atom
  $r(\textbf{d}) = \pi_i(\overline{H_M})$, for a given higher-order Herbrand model, is \emph{true}.
  As a consequence, we may consider $\W$ as a set of possible
  worlds and define this higher-order Herbrand model for $~I:H\rightarrow
  T~$ as a \emph{Kripke model}. It follows that a higher-order language concept $\Box_i A$ is false
 if and only if there is not any possible world where the ground atom
 $A = \pi_i(\overline{H_M}) \in H$ is satisfied, and true if it is satisfied exactly in the
 set of possible worlds that defines the \emph{meaning} of this
 ground atom.\\
  %
We will show, in the following definition, how to define the
accessibility relations for \emph{modal} operators, used to extend
an original many-valued logic by a finite set of higher-order
language concepts. For example,  for any ground modal atom
("concept") $\Box_i A$, where $A = \pi_i(\overline{H_M})$, we will
obtain that $~~| \Box_i A | \in \{ \emptyset, \W \}$, i.e., it is a
\emph{2-valued} modal logic formula (here $\emptyset$ is the empty
set).
  \begin{definition}  \textsc{Kripke semantics for epistemic concepts }: \label{def:Krip} \\ Let $~I:H \rightarrow T~$ be a
higher-order Herbrand interpretation type, where $T $ denotes a
functional space $W_1\Rightarrow (...(W_n \Rightarrow
\textbf{2})...)$, with $~\W = W_1\times...\times W_n$, and $P$ is
the set of predicates in a Herbrand base $H$. Then, for a given
sequence of language concepts $\overline{H_M}$, a quadruple ${\M_I}
= ({\W}, \{{\R}_{i}~|~1\leq i \leq M\}, S, V)$ is a Kripke model
for this interpretation $I$, such that:\\
1. $~S~$ is a non empty set of constants.\\
2. A mapping (see Definition \ref{def:KripSem})$~~V:{\W} \times P
\rightarrow \bigcup_{n \in \N} \textbf{2}^{S^n}$, $~~~~~$ such that
for any $~\textbf{w} = (w_1,...,w_n) \in \W$, $r \in P$ , and
$\textbf{d} \in S^n$ it holds:
$~~~~~~~~~~~~~~V(\textbf{w},r)(\textbf{d})=~I(r(\textbf{d}))(w_1)...(w_n)$,\\
where $S^n$ denotes the set of all n-tuples of constants, and
$\textbf{2}^{S^n}$ the set of all functions from the set $S^n$ to
the set $\textbf{2} $.\\
3. Finite set of accessibility relations: for any $~r(\textbf{d})
 = \pi_i(\overline{H_M})$, $~~~~{\R}_{i} = \W \times Q_i~~$
 if $~Q_i \neq \emptyset$; $~\W \times \W~$ otherwise,
where $~~Q_i = \{\textbf{w}\in \W~|~ V(\textbf{w},r)(\textbf{d})
 =1\}$.\\
 Then, for any world $\textbf{w} \in \W$ and assignment $g$, we define the many-valued satisfaction
 relation, denoted by $~\M_I \models_{g,\textbf{w}} ~$,  as follows:\\
A1.  $\M_I \models_{g,\textbf{w}} ~r(x_1,..., x_n) ~~$ iff $~~V(\textbf{w} ,r)(g(x_1),..., g(x_n)) = 1~~$, for any atom,\\
A2.   $\M_I \models_{g,\textbf{w}} ~\Box_{i}r(x_1,..., x_n) ~~$ iff
$~~\forall \textbf{w'}( ~(\textbf{w}, \textbf{w'}) \in {\R}_{i}~$
  implies $~~\M \models_{g,\textbf{w'}} ~r(x_1,..., x_n))$, \\for any  ground atom  $~r(\textbf{d}) = r(g(x_1),...,
  g(x_n)) \in  \pi_i(\overline{H_M})$.
\end{definition}
Notice that for the introduced higher-order language concepts we
have
that\\
$\M_I \models_{\textbf{w}} ~\Box_{i}r(\textbf{d}) ~$ iff $~\forall
\textbf{w'}( (\textbf{w}, \textbf{w'}) \in {\R}_{i}~$
  implies $~\M \models_{\textbf{w'}} ~r(\textbf{d})) ~$
iff $~\pi_2({\R}_{i}) = | r(\textbf{d})|$.\\
Notice that
we obtained the multi-modal Kripke models with universal modal
operators $\Box_{i}$, that is, we obtained a kind of \emph{2-valued
reduction }for a many-valued atom $~ r(\textbf{d})$. Obviously, this
technique  can only be used  if the number of introduced universal
modal operators is \emph{finite}. \\
 The encapsulated
information in this Kripke frame can be rendered explicit by
flattening a  Kripke model of this more abstract vision of data,
into an ordinary Herbrand model where the original predicates are
extended by set of new attributes for the hidden information.
\begin{definition} \cite{Majk06} \textsc{Flattening}: \label{def:flat}
Let $~I:H\rightarrow T~$ be a higher-order Herbrand interpretation,
where $T $ denotes a functional space $W_1\Rightarrow (...(W_n
\Rightarrow \textbf{2})...)$ and $~\W = W_1\times...\times W_n$ is a
cartesian product. We define its flattening  into the Herbrand
interpretation $I_F:H_F\rightarrow \textbf{2}$, where $~~~~H_F =
\{r_F(\textbf{d},\textbf{w})~|~r(\textbf{d}) \in H $ and
$\textbf{w} \in \W \}$\\
 is the Herbrand base of predicates $r_F$, obtained by an extension  of original predicates $~r~$ by a tuple of parameters $~\textbf{w}
= (w_1,...,w_n)$, such that for any $~~r_F(\textbf{d},\textbf{w})\in
H_F$, it holds that $~~~~I_F(r_F(\textbf{d},\textbf{w})) =
 I(r(\textbf{d}))(w_1)...(w_n)$.
\end{definition}
By this flattening of the higher-order Herbrand models we again
obtain a \emph{2-valued logic}, but with a changed Herbrand base
$H_F$. It can be used as an alternative to the introduction of
universal modal operators, especially when the number of such
operators is \emph{not finite}. Both of these two approaches to the
reduction of many-valued into  2-valued logics will be used in the
rest of this paper, and we will show that the resulting logic in
both cases is a (non truth-functional)  2-valued modal logic.
\section{Contextualization of  many-valued logics} \label{Section:Contextual}

In this Section we will apply the general results obtained in the
previous Section \ref{Section:HiHerbrand} to a more specific case of
many-valued-logics. This is a case of many-valued logics with
uncertain, approximated or context-dependent information.\\
 We consider only the class of many-valued logics
 $\L_{mv}$ based on a bounded lattice $\W$ of algebraic truth
 values, with $\textbf{2} \subset \W$, as explained in the introduction.
   Then the ordering relations and operations in a bounded lattice $\W$
are propagated to the function space $\W^{H}$, that is, to the set
of all Herbrand interpretations, $I_{mv}:H \rightarrow\W$. It is
straightforward ~\cite{Fitt02} that this makes the function space
$\W^{H}$ itself a bounded lattice.
\begin{definition} \label{def:valuation}
Let $\L_{mv}$ be a many-valued logic  with a set of predicate
symbols $P$, a Herbrand base $H$, and with a many-valued Herbrand
interpretation $I_{mv}:H\rightarrow \W$. Then its standard unique
extension to all formulae is a homomorphism $v:\L^G_{mv} \rightarrow
\W$, also called a many-valued \verb"valuation", where $\L^G_{mv}$
is the subset of all ground formulae in $\L_{mv}$. That is, for any
ground formula $X,Y \in \L_{mv}$ holds that\\ $~~~v(\sim X) = \sim
v(X) $ and $~~v(X \circledcirc Y) = v(X)\circledcirc v(Y)$,\\
 where
$\circledcirc $ is any binary many-valued logic connective in
$\Sigma$.
\end{definition}
 Let us, for example, consider  the following bounded
lattices:
\begin{enumerate}
    \item  Fuzzy data ~\cite{Haje98,ZeKa85,DuPr89}: then
    $\W = [0,1]$ is the \emph{infinite} set of real numbers from 0 to 1. For any ground
atom $r(\textbf{d})\in H$ the $p = I(r(\textbf{d}))$ represents its
 \emph{plausibility}. $~$For any two $x,y \in \W$, we have that $x \wedge y = min\{x,y\}$, $~~x \vee y = max\{x,y\}$,
 and negation connective $\sim$ is determined by
$~~$$\sim x =~ 1-x$.\\
     \item Belief quantified data
     ~\cite{KiLi88,NgSu92,LLRS97}:
     then $\W = \C[0,1]$
is the set of all closed subintervals over $[0,1]$. For any ground
atom $r(\textbf{d})\in H$ the $(L,U) = I_{mv}(r(\textbf{d}))$
represents the lower and upper bounds for expert's \emph{belief} in
$r(\textbf{d})$. For any two $[x,y],[x_1,y_1] \in \W$, we have that
$~[x,y] \wedge [x_1,y_1] = [min\{x,x_1\}, min\{y,y_1\} ]$, $~[x,y]
\vee [x_1,y_1] = [max\{x,x_1\}, max\{y,y_1\} ]$.The \emph{belief}
(or truth) ordering  is defined as follows: $~~[x,y] \leq [x_1,y_1]$
$~~$ iff $~~$ $(x \leq x_1$ and $y \leq y_1)$. We define  the
epistemic negation ~\cite{Gins88} of a \emph{belief} $[x,y]$ as the
\emph{doubt} $\sim[x,y]$, such that $\sim[x,y] = [\sim y, \sim x] =
[1-y, 1-x]$. The bottom value of this  lattice is $0 = [0,0]$, while the top value is $1 = [1,1]$.\\
    \item Confidence level quantified data ~\cite{LaSa94,LaSa03}: then $\W =
\C[0,1]\times \C[0,1]$. For any ground atom $r(\textbf{d})\in H$ we
have $((L_1,U_1),(L_2,U_2)) = I_{mv}(r(\textbf{d}))$, where
$(L_1,U_1)$ represents the lower and upper bounds for expert's
\emph{belief} in $r(\textbf{d})$, while $(L_2,U_2)$ represents the
lower and upper bounds for expert's \emph{doubt} in $r(\textbf{d})$,
respectively.\\ Let $\alpha = ([x,y],[z,v])$, $\beta =
([x_1,y_1],[z_1,v_1]) \in
\W$, then,\\
 $~~~\alpha \wedge \beta =~$ $([min\{x,x_1\},min\{y,y_1\}],~[max\{z,z_1\}, max\{v,v_1\}])$,\\
 $~~~\alpha \vee \beta =~ ([max\{x,x_1\},max\{y,y_1\}],~[min\{z,z_1\}, min\{v,v_1\}])$.\\
In this lattice we are interested in the ordering $~\leq~$ that
increases the belief and decreases the doubt of facts,  that is
$~~~$$([x,y],[z,v])~\leq~([x_1,y_1],[z_1,v_1])~$ $~~~$ iff
$~~~~$$[x,y] \leq [x_1,y_1]$ and $[z_1,v_1]\leq [z,v]$.\\
The   negation $~\sim$, which reverses this truth ordering, of this
lattice  is defined by Ginsberg ~\cite{Gins88}, with $~~~~$$\sim
([x,y],[z,v]) = ([z,v],[x,y])$. The bottom value of this  lattice is
$0 = ([0,0],[1,1])$, while the top value
is $1 = ([1,1],[0,0])$.\\
\item Belnap's bilattice  based logic programs ~\cite{Fitt91}:
Then its truth lattice is $\W = \B = \{f,t,\top,\bot\}$, where $1 =
t$ is \emph{true}, $0 = f$ is \emph{false}, $\top$ is inconsistent
(both true and false) or \emph{possible}, and $\bot$ is
\emph{unknown}. As Belnap observed, these values can be given a
\emph{truth} ordering, $\leq_t$, such that $0 \leq_t \top \leq_t 1$,
$~0 \leq_t \bot \leq_t 1$ and $\bot \bowtie_t \top$,  with $\alpha
\wedge \beta = min_t\{\alpha , \beta\},~\alpha \vee \beta =
max_t\{\alpha , \beta\}$, and the epistemic negation $\sim$ is
defined by: $\sim 0 = 1, ~\sim 1 = 0, \sim \bot = \bot, ~\sim \top =
\top$.
\end{enumerate}
All examples above are more than bounded lattices: they are complete
distributive lattices \cite{Majk06Bi,Majk07MV}. Thus, we consider
also that for any two elements $a,b \in \W$ the many-valued
implication $a\rightarrow b$ for complete lattices can be defined as
a reduct (the relative pseudocomplement), that is $a\rightarrow b =
\vee \{c \in \W~|~c \wedge a \leq b\}$, so that $a\rightarrow b = 1$
iff $1
\wedge a = a \leq b$. \\
 For a given \emph{many-valued} logic $\L_{mv}$, we can
generate a \emph{contextual} logic $\L_{ct}$, so that for any ground
atom $r(\textbf{d}) \in H$ with a logic value $\textbf{w} =
I_{mv}(r(\textbf{d})) $, we generate a \emph{contextual atom}, a
couple $(r(\textbf{d}) , \textbf{w})\in H\times \W$, which tell us
that "the atom $r(\textbf{d})$ in the context $\textbf{w}$ is true".
We also define  the extended Herbrand base $~~~~H_F =
\{r_F(\textbf{d},\textbf{w})~|~r(\textbf{d}) \in H $ and $\textbf{w}
\in \W \}$ by extending each original atom by the logic attribute
with the domain $\W$, and with the bijection $is:H_F \rightarrow
H\times \W$, such that
for any extended (or flattened) ground atom $r_F(\textbf{d},\textbf{w}) \in H_F$ it holds that
 $is(r_F(\textbf{d},\textbf{w})) = (r(\textbf{d}), \textbf{w})$. \\
This \emph{contextualization} of a many-valued logic can be
represented by the following commutative diagram $\vspace*{-3mm}$
\begin{diagram}
  && \textbf{2}^{\W}\times \W  &   & \rTo^{eval}& & \textbf{2} \\
\verb"Higher-order"~~~~ & & \uTo^{I} \uTo^{~~id_W}&  &  &  &\uTo_{id_\textbf{2}} \\
&&H \times \W  &          \lTo_{is} & H_F & \rTo_{I_F}  &\textbf{2}  \\
\verb"Many-valued"~~~~& ~~~~& \dTo^{I_{mv}} \dTo^{~~id_W}&   & & & \uTo_{id_\textbf{2}}\\
& &\W \times \W  &      & \rTo^{\bigtriangleup} & &  \textbf{2}
\end{diagram}
where $eval$ is the application of the first argument (function) to
the second argument, $id$'s are the identities, and $\bigtriangleup$
is the 'diagonal' function, such that
$~~~~\bigtriangleup(\textbf{w},\textbf{w}') = 1~~$ iff $~~\textbf{w}
= \textbf{w}'$, so that the higher-order Herbrand interpretation is
obtained from a many-valued Herbrand interpretation by $I =
[\bigtriangleup \circ (I_{mv} \times id_W)]$, where $~[\_]$ is the
currying ($\lambda$ abstraction) operator for functions. The
flattened Herbrand interpretation (of a 'meta' logic obtained by an
ontological encapsulation of original many-valued logic), is equal
to: $~~I_F = eval \circ ([\bigtriangleup \circ (I_{mv} \times id_W)]
\times id_W) \circ is$.\\
Intuitively, the diagram above shows that for any many-valued
interpretation $I_{mv}$, we obtain the correspondent 2-valued
interpretation $I_F$ (but with modified Herbrand base $H_F$), and,
equivalent to it, the higher-order Herbrand interpretation $I$. \\
By this contextualization of a many-valued logic we obtain the
simplest case of the higher-order Herbrand interpretation given by
Definition \ref{def:HOHerbrand}, $~~~I:H \rightarrow
\textbf{2}^{\W}~$, such that for any atom $r(\textbf{d})\in H$ and
$\textbf{w}\in \W$ holds that:\\
   $~~~~~~~~I(r(\textbf{d}))(\textbf{w}) = 1,$ $~~$ iff $~~$
$\textbf{w} = I_{mv}(r(\textbf{d}))$.\\
The accessibility relations $~~{\R}_{i} = \W \times Q_{i}$, for any
$~r(\textbf{d}) = \pi_i(\overline{H_M}) \in H$, in Definition
\ref{def:Krip} for many-valued logic does not depend on the number
of ground atoms in a Herbrand base, but only on the number of logic
values in $\W$:
 it results from the fact that to any ground atom in a \emph{consistent} many-valued
logic we can assign only \emph{one} logic value, so that
 $~~Q_{i} = \{\textbf{w} \in \W~|~r(\textbf{d}) = \pi_i(\overline{H_M}) \in H$ and $~I(r(\textbf{d}))(\textbf{w})
  =1\} = \{ \textbf{w} \} $ is a singleton, with $\textbf{w} = I_{mv}(r(\textbf{d}))$.\\
  Thus,  we are able to make the reduction to
  2-valued logic by the introduction of a  number of universal
  modal operators $ \Box_{\textbf{w}}$ (denoted also by $[ \textbf{w}]$ in what follows) with the
  accessibility relation $~~{\R}_{\textbf{w}} =
\W \times Q_{\textbf{w}} = \W \times \{ \textbf{w}\} $, for each
$~\textbf{w} \in \W$.\\
Each universal modal operator $[\textbf{w}]$, with the meaning " has
the value $\textbf{w}$", is defined algebraically in a lattice $\W$
as a unary operator (function) $[\textbf{w}]:\W \rightarrow
\textbf{2} \subseteq \W$, such that for any $\textbf{w}_1 \in \W$,
$~~~[\textbf{w}](\textbf{w}_1) = 1$ if $\textbf{w}_1 = \textbf{w}$;
$~~0$ otherwise.
\\ These modal operators \emph{are not monotonic}
 operators, so that
we obtain a non-normal Kripke modal logic (for example, the necessity rule does not hold). \\
As we can see, we assume that the set of possible worlds of the
relational Kripke frames, used for the transformation of many-valued
into multi-modal 2-valued logic, is the set of logic values of this
many-valued logic. This is an autoreferential semantics
\cite{Majk06ml,MaPr09} and a formal result of the modal
transformation for higher-order Herbrand models and the
transformation of many-valued Herbrand models into higher-order
Herbrand models. The philosophical assumption is, instead, that each
possible world represents a level of \emph{credibility}, so that
only the propositions with the right logic value (i.e., level of
credibility) can be accepted by this world.
\section{Reduction of many-valued into 2-valued multi-modal  Logic Programs \label{Section:LogProg}}
  Let $PR$ be a many-valued logic program, for a given many-valued logic $\L_{mv}$ with a set of algebraic truth-values given by a bounded
  lattice $\W$, a Herbrand base $H$ and a many-valued Herbrand interpretation $I_{mv}:H \rightarrow
  \W$ that is also a \emph{model} of $PR$, i.e., an interpretation that satisfies all logic
  clauses in a logic program $PR$. We denote by $Mod$ the subset of
  all Herbrand interpretations in $\W^H$ that are also models of $PR$. Then we will have the  following two
cases: \\
1. In the first case by introducing the set of unary modal operators
for each algebraic logic value in $\W$ (both for finite and infinite
cases) we obtain the \emph{standard} 2-valued modal logic for the
satisfaction of logic conjunction and disjunction (if a proposition
is defined by the set of worlds where it is satisfied, then the
conjunction/disjunction of any two propositions is equal to the set
intersection/union respectively), by transforming many-valued ground
atoms into 2-valued \emph{modal} ground atoms.\\
2. In the second case we do not use one specific unary modal
operator for each given algebraic logic value, which can be somewhat
complex issue when the cardinality of $\W$ is very big or infinite.
We do not transform the many-valued logic connectives into the
standard 2-valued logic connectives as in the first case: instead,
they will be transformed into \emph{binary} modal operators with the
ternary accessibility relations . In order to obtain a non standard
modal logic in which
  the intersection/union properties hold for conjunction/disjunction
respectively, we also need to introduce an existential modal
operator with binary accessibility relation equal to the cartesian
product of possible worlds. The semantics of this approach is more
complex and transforms all original atoms of the many-valued logic,
but offers one advantage because the number of modal operators is
small, equal to the number of  logic operators in the original
many-valued logic.
\subsection{Unary modal operators case}
%
 We will show
how a many-valued logic program can be transformed into the 2-valued
multi-modal logic
program \emph{without} modifying the original set of atoms of a many-valued logic program.\\
As we have seen, by the contextualization of a many-valued logic
$\L_{mv}$ we obtain a contextual logic $\L_{ct}$ with the same
Herbrand base $H$ as the original many-valued logic but (for a given
many-valued Herbrand model $I_{mv} \in Mod$) with a higher-order
model $I = [\bigtriangleup \circ (I_{mv} \times id_W)]:H \rightarrow
\textbf{2}^{\W}$ as has been shown by the commutative diagram in
Section \ref{Section:Contextual}.
 We are now able  to apply the result of the method in Definition \ref{def:Krip}
to this contextual logic with  higher-order model types.\\
    A simple modal formulae
$[w]p(x_1,..,x_n)$, where $w \in  \W$ and $p(x_1,..,x_n) $ is an
atom of the many-valued logic program $PR$, will be called m-atom
(\emph{modal atom}). A  2-valued multi-modal logic, obtained by the
substitution of original many-valued atoms by these m-atoms, is
considered the first time in the case of the 4-valued Belnap's
logics, used for databases with incomplete and inconsistent
information ~\cite{MajkC04}.
 \begin{definition}(Program Transformation: Syntax)  \label{def:transorm}
 Let $PR$ be a many-valued  lattice-based logic program.
  We define its transformation  in the correspondent
positive multi-modal logic program $P_{mm}$ as follows (bold
constants and variables denote tuples):
 \\ 1. Each ground atom in the original many-valued program $PR$, $~~~~p(\textbf{c}) \leftarrow ~\alpha$,\\
where  $\alpha \in \W$ is a fixed logic value, we transform into the
following  2-valued ground m-atom clause in $P_{mm}$:
$~~~(1)~~~~[\alpha]p(\textbf{c}) \leftarrow ~~~~$\\
2. Each set of original many-valued clauses in $PR$, with the same
head, (here $\vee, \wedge $ are a many-valued disjunction and
conjunction respectively, i.e., the join and meet operators of a
lattice $\W$, and
$S$ is a finite interval of natural numbers from 1 to $n$), \\
 $~~~~p(\textbf{x})
~\leftarrow ~\vee_{ j \in S}(
~r_{j,1}(\textbf{x}_{j,1}),...,r_{j,k_j}(\textbf{x}_{j,k_j}), \sim r_{j,k_j+1}(\textbf{x}_{j,k_j+1}),..., \sim r_{j,m_j}(\textbf{x}_{j,m_j}))$,\\
we transform as follows: \\ let us denote by $Var_w= \bigcup_{ j \in
S} \{v_{j,1},..., v_{j,k_j}, v_{j,k_j+1},...,v_{j,m_j}\}$ the set of
logic variables for atoms in this clause. Then, for each assignment
$g:Var_w \rightarrow \W$ we define a new 2-valued clause with
m-atoms, and with the classic 2-valued
disjunction $\bigvee$, in  $P_{mm}$ :\\
$(2)~~~ [\beta]p(\textbf{x}) ~\leftarrow ~\bigvee_{ j \in S}(
~[\alpha_{j,1}]r_{j,1}(\textbf{x}_{j,1}),...,[\alpha_{j,k_j}]r_{j,k_j}(\textbf{x}_{j,k_j}),$ \\
$,
~~~~~~~~~~~~~~~~~~~~~~~~~~~~~~~~~~~~~~[\alpha_{j,k_j+1}] r_{j,k_j+1}(\textbf{x}_{j,k_j+1}),..., [\alpha_{j,m_j}] r_{j,m_j}(\textbf{x}_{j,m_j}))$,\\
where $\alpha_{j,i} = g(v_{j,i})$, for  $j \in S$, $1\leq i \leq m_j$, and\\
$ \beta = ~\vee_{ j \in S}(g(v_{j,1}) \wedge ...\wedge g(v_{j,k_j})
\wedge \sim g(v_{j,k_j+1})\wedge ...\wedge \sim g(
v_{j,m_j}))$.\\
 Consequently, based on clauses (1) and (2), we obtain a
standard positive logic program $P_{mm}$ with 2-valued m-atoms.
\end{definition}
\textbf{Remark}: notice that the obtained positive multi-modal logic
program $P_{mm}$ uses only standard 2-valued logic connectives, in
contrast to the original many-valued logic program $PR$ where the
logic connectives are lattice-based (many-valued) logic operators.
The grounded program $P^G_{mm}$ obtained from the program $P_{mm}$,
by substituting in all possible ways the variables of its atoms in
all its clauses, will contain only 'modal ground atoms'
$[\alpha_{k_j}]r_{j,k_j}(\textbf{d}_{j,k_j})$. To such atomic
formulae we can assign the new fresh propositional symbols, so that
with these propositional symbols the program $P^G_{mm}$ becomes a pure 2-valued logic program. \\
 As we can verify, the
obtained multi-modal logic program $P_{mm}$ is a \emph{positive}
logic program (without negation in the body of clauses), so that it
has a \emph{unique} model (the set of all true facts derivable from
this 2-valued logic program).
\begin{propo} (Invariance) For any given many-valued logic
program $PR$, the transformed 2-valued logic program $P_{mm}$  with
modal atoms has the same Herbrand model $I_{mv}:H\rightarrow \W$ as
the original program $PR$.
\end{propo}
\textbf{Proof:} We have to show that for a given logic program $PR$,
its many-valued Herbrand model $I_{mv}:H\rightarrow \W$ also
satisfies  the clauses of the positive 2-valued modal program
$P_{mm}$. We will consider their grounded versions, $PR_G$ and
$P^G_{mm}$ respectively. Then, for any ground fact $~p(\textbf{c})
\leftarrow ~\alpha$ we have that $~I_{mv}(p(\textbf{c})) = \alpha$,
so that for the modal operator $[\alpha]:\W \rightarrow \textbf{2}$,
$~~[\alpha](I_{mv}(p(\textbf{c})))~= [\alpha](\alpha) = 1~~$ and,
consequently, the correspondent modal fact in $P^G_{mm}$,
$~~[\alpha]p(\textbf{c}) \leftarrow ~$, is satisfied by $I_{mv}$.\\
Let us consider a ground clause in $PR_G$,\\
$(1)~~p(\textbf{c}) ~\leftarrow ~\vee_{ j \in S}(
~r_{j,1}(\textbf{c}_{j,1}),...,r_{j,k_j}(\textbf{c}_{j,k_j}), \sim r_{j,k_j+1}(\textbf{c}_{j,k_j+1}),..., \sim r_{j,m_j}(\textbf{c}_{j,m_j}))$,\\
which is satisfied by the model $I_{mv}$ with logic values $w =
I_{mv}(p(\textbf{c})) $ and $w_{j,i_j} =
I_{mv}(r_{j,i_j}(\textbf{c}_{j,i_j}))$ for $1\leq j \leq n$ and
$1\leq i \leq m$, such that $ w = ~\vee_{j \in S}(w_{j,1} \wedge
...\wedge w_{j,k_j} \wedge \sim w_{j,k_j+1}\wedge ...\wedge \sim
w_{j,m_j})$.  Then, the transformation of this ground clause (1) of
$PR_G$ into the 2-valued modal
clauses will be the following \emph{set} of modal rules\\
$(2)~~~ [\beta]p(\textbf{c}) ~\leftarrow ~\bigvee_{ j \in S}(
~[\alpha_{j,1}]r_{j,1}(\textbf{c}_{j,1}),...,[\alpha_{j,k_j}]r_{j,k_j}(\textbf{c}_{j,k_j}),$ \\
$,
~~~~~~~~~~~~~~~~~~~~~~~~~~~~~~~~~~~~~~[\alpha_{j,k_j+1}] r_{j,k_j+1}(\textbf{c}_{j,k_j+1}),..., [\alpha_{j,m_j}] r_{j,m_j}(\textbf{c}_{j,m_j}))$,\\
for \emph{all combinations} of $\alpha_{j,i}  \in \W$, for $j \in S$, $1\leq i \leq m_j$, and\\
$ \beta = ~\vee_{j \in S}(\alpha_{j,1}  \wedge ...\wedge
\alpha_{j,k_j} \wedge \sim \alpha_{j,k_j+1}\wedge ...\wedge \sim
\alpha_{j,m_j})$.\\
It is easy to verify that the body of the ground rule (2) is true
only iff $\alpha_{j,i_j} = w_{j,i_j}$ for all $j \in S$ and $1\leq i
\leq m$, and that in that case $\beta = w$, so that
$[\beta](I_{mv}p(\textbf{c})) = [\beta](w) = [\beta](\beta) = 1$,
that is, also the head of this rule is true, so that this clause is
satisfied. For any other combination of modal operators in every
other rule (2), derived from the rule (1),  we obtain that its body
is false, thus such a rule in $P_{mm}$ is satisfied by $I_{mv}$.
Consequently, the Herbrand model $I_{mv}$ also satisfies  the
transformed 2-valued logic program $P_{mm}$. From the fact that
$P_{mm}$, as a positive logic program, can have only one Herbrand
model we conclude that $I_{mv}$ is the \emph{unique} Herbrand model
of $P_{mm}$, and thus the program transformation is correct and
knowledge invariant.
\\ $\square $ \\
Now we will consider a Kripke model for this transformed modal
2-valued logic program $P_{mm}$ based on  Definition \ref{def:Krip},
based on the particular accessibility relations for introduced unary
modal operators previously discussed.
\begin{definition}(Program Transformation: Semantics)  \label{def:transSem}
\\ Let $PR$ be a many-valued  lattice-based logic program with
  a many-valued Herbrand model $I_{mv}:H\rightarrow \W$, where $H$ is a Herbrand base with a set $P$ of predicate symbols.
 Its correspondent positive multi-modal
logic program $P_{mm}$ has the Kripke model ${\M_I} = ({\W},
\{{\R}_w~|~w \in \W\}, S, V)$, with ${\R}_w = \W \times \{ w\} $ and
 $~~V:\W\times P \rightarrow {\bigcup}_{n \in \N} \textbf{2}^{S^n}$ (from Definition \ref{def:KripSem}), such that for any $p \in P$ with arity n, a set
of constants $ (c_1,..,c_n) \in S^n $, and a world $w \in \W$,
$~~V(w, p)(c_1,..,c_n) = 1~~$ iff $~~w =
I_{mv}(p(c_1,..,c_n))$.\\
Then, for any assignment   $g $ and $w \in \W$,  the  satisfaction relation $\models_{g,w}$ is defined as follows: \\
1. $~\M_I\models_{g,w} p(x_1,..x_n)~~$ iff
$~~V(w,p)(g(x_1),..,g(x_n)) = 1$.\\
2. $~~\M_I\models_{g,w} [\alpha]p(x_1,..x_n) ~~$ iff $~~\forall
y((w,y)\in \R_{\alpha}$
implies $~\M_I\models_{g,y} p(x_1,..x_n))$. \\
3. $~\M_I\models_{g,w} \phi \bigwedge \psi ~~$ iff
$~~\M_I\models_{g,w}
\phi ~$ and $~\M_I\models_{g,w} \psi $.\\
4. $~\M_I\models_{g,w} \phi \bigvee \psi ~~$ iff
$~~\M_I\models_{g,w}
\phi ~$ or $~\M_I\models_{g,w} \psi $.\\
5. $~\M_I\models_{g,w} \phi \rightarrow \psi ~~$ iff
$~~\M_I\models_{g,w} \phi ~$ implies $~\M_I\models_{g,w} \psi $.
\end{definition}
\textbf{Remark}: We obtained a  modal logic for the multi-modal
program $P_{mm}$ in Definition \ref{def:transorm}. If we denote by
$|\psi/g|$ the set of worlds where the ground formula $\psi/g$ is
satisfied, then
 $~|p(g(x_1),..,g(x_n))|$  is a singleton set.\\
Thus, differently from the original ground atoms that can be
satisfied in a singleton set only,  the modal atoms have a standard
2-value property, that is, they are true or false in the Kripke
model, and consequently are satisfiable in all possible worlds, or
absolutely not satisfiable in any world. Consequently, our positive
modal program with modal atoms satisfies the classic 2-valued
properties:
\begin{propo} \label{Prop:2-valued}For any ground formula $\phi/g$ of a positive multi-modal
logic program $P_{mm}$ in Definition \ref{def:transorm}, we have
that $|\phi/g| \in \{\emptyset, \W\}$, where $\emptyset$ is the
empty set.
\end{propo}
\textbf{Proof}: by structural induction :\\
1. $~|[\alpha]p(x_1,..x_n)/g| = \W~~$ if $~\alpha =
I_{mv}(p(g(x_1),..,g(x_n)))$; $~~\emptyset$, otherwise.\\
Let, by inductive hypothesis, $|\phi/g|, |\psi/g| \in \{\emptyset,
\W\}$, then\\
2. $~\M_I\models_{g,w} \phi \bigwedge \psi ~~$ iff $~w \in |(\psi \bigwedge \phi)/g| = |\psi/g| \bigcap |\psi/g| \in \{\emptyset, \W\}$.\\
3. $~\M_I\models_{g,w} \phi \bigvee \psi ~~$ iff $~w \in |(\psi \bigvee \phi)/g| = |\psi/g| \bigcup |\psi/g| \in \{\emptyset, \W\}$.\\
4. $~\M_I\models_{g,w} \phi \rightarrow \psi ~~$ iff $~~w \in
|\phi/g|$ implies $~w \in |\psi/g|$. Thus, $\phi \rightarrow \psi$
is true in the model $\M_I$ iff $~~ |\phi/g| \subseteq |\psi/g|$,
that is, iff $~ v_B(\phi/g) \leq v_B(\psi/g)~$, or, alternatively,
$~ \phi/g \vdash \psi/g~$,
where $\vdash$ is the deductive inference relation for this 2-valued modal logic.\\
Thus, $|\phi/g \rightarrow \psi/g| = (\W - |\phi/g |) \bigcup
|\psi/g| \in \{\emptyset, \W\}$.
\\ $\square$ \\
 The following proposition demonstrates the existence of a
one-to-one correspondence between the unique many-valued model of
the original many-valued logic program $P$ and this unique
multi-modal positive logic program $P_{mm}$.
\begin{propo} Let $PR$ be a many-valued logic program with a
Herbrand model $I_{mv}:H \rightarrow \W$, then the model $\M_I$ of
the 2-valued multi-modal program $P_{mm}$, obtained by the
transformation defined in Definition \ref{def:transorm}, is composed
by the set of true atomic formulae  $~~~~S_T =
\{[\alpha]p(\textbf{c})~|~p(\textbf{c}) \in H$ and $\alpha =
I_{mv}(p(\textbf{c})) \}$.
\end{propo}
\textbf{Proof:} For any ground atom $p(\textbf{c}) \in H$ such that
its logic value in a Herbrand model $I_{mv}:H \rightarrow \W$,
obtained by Clark's completion ~\cite{Fitt85,Kune87,Geld87},  is
equal to $\alpha = I_{mv}(p(\textbf{c}))$, we have that
$~~|[\alpha]p(\textbf{c})| = \{w \in \W~|~\M_I\models_{w}
[\alpha]p(\textbf{c})\} = \{w \in \W~|~\alpha =
I_{mv}(p(\textbf{c}))\} = \W$.
Thus, $~~~~[\alpha]p(\textbf{c})$ is true in  $~\M_I~$.\\
$\square$\\
\textbf{Remark:} This transformation of many-valued logic programs
into \emph{positive} (without negation) logic programs (but with
\emph{modal} atoms), can also be used  to manage the
\emph{inconsistency} in 2-valued logic programs: while in the
original 2-valued logic we are not able to manage the ground atom
$p(c_1,..c_n)$ that is both true and false without the explosive
inconsistency of all logic, in the transformed positive modal
program we can have the ground modal atoms $[1]p(c_1,..c_n)$ and
$[0]p(c_1,..c_n)$ both true without generating the inconsistency.
This means that
this kind of  2-valued transformation can be used for \emph{paraconsistent} logics, as shown in the example below.\\
 \textbf{Example 1}: The
smallest \emph{nontrivial} bilattice is Belnap's 4-valued bilattice
~\cite{Beln77,Fitt91} $\W = \B = \{f,t,\bot, \top\}$where $t$ is
\emph{true}, $f$ is \emph{false}, $\top$ is inconsistent (both true
and false) or \emph{possible }, and $\perp$ is \emph{unknown}. As
Belnap observed, these values can be given two natural orders:
\emph{truth} order, $\leq_t$, and \emph{knowledge} order, $\leq_k$,
such that $f \leq_t \top \leq_t t$, $~f \leq_t \bot \leq_t t$, $\bot
\bowtie_t \top$ and $\bot \leq_k f \leq_k \top$, $~\bot \leq_k t
\leq_k \top$, $f\bowtie_k t$. That is, the bottom element $0$ for
$\leq_t$ ordering is $f$, and for $\leq_k$ ordering is $\bot$,
 and the top element $1$ for $\leq_t$ ordering is $t$, and for $\leq_k$ ordering is $\top$\\
Meet and join operators under $\leq_t$ are denoted $\wedge$ and
$\vee$; they are natural generalizations of the usual conjunction
and disjunction notions. Meet and join under $\leq_k$ are denoted
$\otimes$ and $\oplus$, such that it holds that: $~f \otimes t =
\bot$, $f \oplus t =\top$, $\top\wedge \bot
= f$ and $\top \vee \bot = t$.\\
We may use  a \emph{relative pseudo-complements} for the
implication, defined by $x \rightharpoonup y = \vee\{z~|~z \wedge x
\leq_t y\}$, and the \emph{pseudo-complements} for the negation,
 $\neg_t x = x \rightharpoonup f$.\\
 In Belnap's bilattice the conflation $-$ is a monotone function
 that
preserves all finite meets (and joins) w.r.t. the lattice $(\B,
\leq_t)$, thus it is the universal (and existential, because $- =
\neg_t - \neg_t$) modal many-valued operator: " it is believed
that", which extends the 2-valued belief of the autoepistemic
logic as follows:\\
1. if $A$ is true than "it is believed that A", i.e., $-A$, is
  true;\\
 2. if $A$ is false than "it is believed that A" is
  false;\\
3. if $A$ is unknown than "it is believed that A" is
  inconsistent: it is really inconsistent to believe in something
  that is unknown;\\
 4.  if $A$ is inconsistent (that is, \emph{both} true and false) then "it is believed that A" is
  unknown: really, we can not tell anything about belief in something that is
  inconsistent.\\
  This belief modal operator is used to define the  \emph{epistemic negation} $\neg$, as composition of
the strong negation $\neg_t$ and this belief operator, i.e.,  $\neg
=
\neg_t -$.\\
Let us show how these modal atoms  in Definitions \ref{def:transorm}
and \ref{def:transSem} can be used for paraconsistent logic, able to
deal with the truth of the formulae $B = A \wedge \neg A$ as well:
when a formula $B$ is \emph{true} then a formula $A$ is called
\emph{inconsistent}, that is, has the logic value $\top$ in the
Belnap's 4-valued logic. It is easy to see that in such a case a
formula $B$ corresponds to a 2-valued formula $[\top]A$, i.e.,
$[\top]A = A \wedge \neg A$, where the modal operator $[\top]$ is an
"it is inconsistent" operator (used also as $\bullet$ in Logics of
Formal Inconsistency (LFI) \cite{AmCM02} for the 3-valued sublattice
$\B_3 = \{f, \top, t\} \subset
\B_4$).\\
But the other operator $[\bot]$ is a modal "it is \emph{unknown}"
 operator, used to support  an incomplete knowledge as well. That
is, when a formula $[\bot]A$ is true, then a formula $A$ is called
\emph{unknown}, and has the value $\bot$ in Belnap's 4-valued logic.\\
This is the reason why we are using  Belnap's 4-valued logic for the
paraconsistent data integration \cite{MajkA04} of partially
inconsistent and  incomplete information. In \cite{MajkA04} we use
 the 4-valued logic directly with  Moore's \emph{autoepistemic}
operator \cite{Gins88},  $~\mu:\B \rightarrow \B$,  for a Belnap's
bilattice, defined by $\mu(x) = t$ if
$x \in \{\top, t\}$; $~~f~~$ otherwise.\\
It is easy to verify that it is monotone w.r.t.  $\leq_t$, that is,
it is multiplicative ($\mu(x \wedge y) = \mu(x) \wedge \mu(u)$ and
$\mu(t) = t$) and additive ($\mu(x \vee y) = \mu(x) \vee \mu(u)$ and
$\mu(f) = f$). Consequently,  it is a selfadjoint (contemporary
universal and existential) modal operator, $\mu = \neg_t \mu
\neg_t$. But if we are adopting, alternatively,  the proposed
2-valued reduction for this Belnap's 4-valued logic, we are able to
use the modal operators $[\bot]$ and $[\top]$ in order to deal with
incomplete and inconsistent information as well.
%
\subsection{Binary modal operators case}
%
In this subsection we will  use an alternative method w.r.t. the
precedent case, based on a flattening, in order to reduce a
many-valued into a 2-valued logic.
 The flattening of an original many-valued
lattice-based program into a modal meta logic is a kind of
\emph{ontological-encapsulation}, where the encapsulation of an
original many-valued logic program into the 2-valued modal meta
logic program corresponds to a flattening process described in
Definition \ref{def:flat}.
This approach is developed in a number of papers, and more
information can be found in \cite{Majk04on,MajkC04,Majk05f,Majk06}.
Here we will present a slightly modified version of this ontological
encapsulation.\\
We will also
introduce a new symbol $~\textbf{e}~$ (for "error condition"),
necessary in order to render \emph{complete} the functions for a
generalized interpretation and a semantic-reflection, defined w.r.t.
a particular model $I_{mv} \in Mod$, as  follows:
\begin{definition} \label{def:reflex}
Let $PR$ be a many-valued logic program with a set of predicate and
functional symbols $P$ and $F$ respectively, with a Herbrand model
$I_{mv}:H\rightarrow \W$ where $H$ is a Herbrand base,  with a set
$\T_0$ of all ground terms and a set $\T = \bigcup_{k \in
\N}\T_0^k$ with $\N = \{1,2,...,n\}$ where $n$ is the maximal arity of symbols in $P \bigcup F$.\\
A generalized interpretation  is a mapping $~\I:P \times \T
\rightarrow \W \bigcup \{\textbf{e}\}$, such that for any
$\textbf{c} = (c_1,..,c_n) \in \T $,
 $~~~\I(p, \textbf{c}) = I_{mv}(p(\textbf{c}))~~$ if $~~
ar(p) = n$; $~~\textbf{e}~$ otherwise.\\
 Then, a semantic-reflection  is defined by a mapping $~\K =\lambda \I:P
\rightarrow (\W \bigcup \{\textbf{e}\})^{\T}$, where $\lambda$ is
the currying operator from lambda calculus. \\ For each $~p\in \P$
that is not a built-in 2-valued predicate, we define a new
functional symbol $~\kappa_p~$ for a mapping $~
\K(p):\T \rightarrow \W \bigcup \{\textbf{e}\}$.\\
If $p$ is a 2-valued built-in predicate, then  the mapping
$\kappa_p$ is
 defined uniquely and independently of $I_{mv}$, by: for any $\textbf{c} \in \T_0^{ar(p)}$, $~~\kappa_p(\textbf{c}) = 1$ if
$~p(\textbf{c})$ is true; $0~$ otherwise.
\end{definition}
We recall the well-known fact that 2-valued \emph{built-in}
predicates (as $\leq$, $=$, etc..) have  constant extensions in any
Herbrand interpretation (they preserve \emph{the same meaning}
for any logic interpretation, differently from ordinary predicates).\\
 A semantic-reflection $\K$, obtained from a generalized
interpretation $\I$, introduces a function symbol $\kappa_p = \K(p)$
for each predicate $p \in P$ of the original logic program $PR$,
such that for any $\textbf{c} = (c_1,..,c_n) \in \T $, it holds that
$~~~\kappa_p(\textbf{c}) = I_{mv}(p(\textbf{c}))~$ if $~ ar(p) = n$;
$\{\textbf{e}\}$ otherwise. These new function symbols will be used
in a new meta logic language, used to transform each original
many-valued atom $p$ in $P$ into a new  atom $p_F$ obtained as an
extension of the original atom $p$ by one "logic" attribute with the
domain of values in $\W$. The interpretation of a function symbol
$\kappa_p$ in this new meta logic program has to reflect the meaning
of the original many-valued predicate $p$ in the original
many-valued logic program $PR$.
This is the main reason why we are using the name
\emph{semantic-reflection} for a mapping $\K$, because by
introducing the many-valued interpretations contained in the set of
built-in functional symbols $~\kappa_p~$ as objects of a meta logic
(defined in the following Definition \ref{def:flatt}),
 the obtained  logic becomes a \emph{meta-logic} w.r.t. the original many-valued logic.
 Consequently, we are able to introduce
  a program encapsulation (flattening)
transformation $\E$, similarly as in ~\cite{Majk04on}, as follows:
\begin{definition} \label{def:flatt}(Ontological encapsulation of Many-valued Logic Programs: Syntax)\\ Let $PR$ be a many-valued logic
program with a set of predicate symbols $P$, a many-valued Herbrand
model $I_{mv}:H\rightarrow \W$, and a semantic-reflection $\K $.
  Then, the translation $\E$ of a program $PR$ into
its encapsulated syntax version  $PR_F$ is as follows: for each predicate symbol $p \in P$ with arity $n$, we introduce a predicate symbol
$p_F$ with one more attribute with a domain in $\W$. Then,\\
1. each atom  $p(t_1,..,t_n))$ in $PR$ with terms $t_1,...,t_n$, we transform as follows \\
$~~~\E(p(t_1,..,t_n))
= p_F(t_1,..,t_n,\kappa_p(t_1,..,t_n))$, \\
and we denote by $P_F$ the set of all new obtained predicates
$p_F$.\\
For any formula $\phi, \varphi \in \L_{mv}$ we do as follows:\\
2. $\E(\sim \phi) =  ~\sim ^A \E(\phi)$; \\
 3. $\E(\phi \wedge \varphi) = \E(\phi) \wedge^A \E(\varphi)$;
$~~\E(\phi \vee \varphi) = \E(\phi) \vee \E(\varphi)$;
 $~~\E(\phi \leftarrow \varphi) = \E(\phi) \leftarrow^A \E(\varphi)$,\\
 where $\wedge^A, \vee^A$ and $\leftarrow^A$) are new introduced binary symbols for the conjunction, disjunction and implication, at the
encapsulated  meta level, respectively. Thus, the obtained meta
program  $PR_F = \{\E(\phi)~|~ \phi $ is a clause in $PR \}$, has a
Herbrand base $~~~H_F = \{~p_F(c_1,..,c_n,\alpha)~|~p(c_1,..,c_n)
\in H $ and $ \alpha \in
\W \}$.\\
 We denote by $\L_F$ the set of
formulae (free algebra) obtained from the  set of predicate letters
in $P_F$ and modal operators $\sim^A, \wedge^A, \vee^A$ and
$\leftarrow^A$.
\end{definition}
\textbf{Remark:} the new introduced logic symbols $\sim^A, \wedge^A,
\vee^A$ and $\leftarrow^A$ for the metalogic operators of negation,
conjunction, disjunction and implication are not necessarily
truth-functional as are original many-valued operators ($\wedge$ and
$\leftarrow$ for example) but rather are modal (non
truth-functional). The unary operator $\sim^A$ is not a negation
(antitonic) operator but a modal operator, so that by this
transformation of $PR$  we obtain a modal logic program $RP_F$ that
is a \emph{positive} logic program (without negation). Differently
from a ground many-valued formula $\phi \in \L_{mv}$, the
transformed meta-formula $~\E(\phi) \in \L_F$ can be only \emph{true
or false} in a given possible world $w \in \W$ for this meta modal
logic (in a given Kripke model $\M$ of obtained meta logic program
$PR_F$).
\\
In this definition of a meta logic program $PR_F$, the set of
mappings $\{\kappa_p = \K(p)~|~p \in P_S \}$ is considered as a set
of \emph{built-in functions}, determined by a given semantic
reflection $\K$,  that extends a given set of
functional symbols in $F$.\\
 This embedding of a many-valued logic program $P$ into a
 meta logic program $P_F$ is an \emph{ontological}
embedding: it considers both the formulae of $PR$  with their
many-valued interpretation obtained by semantic reflection (a set of
built-in functions $\kappa_p$) of original many-valued logic in this
new modal meta logic.
\\
The encapsulation operator $\E$ is intended to have the following
property for  a valuation $v$ (a homomorphic extension of Herbrand
interpretation $I_{mv}$ to all formulae in $\L_{mv}$ given by
Definition \ref{def:valuation}) of a many-valued logic program
$PR$:\\ for  any ground many-valued formula $\phi$, the encapsulated
meta formula $\E(\phi)$ intends to capture the notion of $\phi$ with
its value $v(\phi)$ as well, in the way that "$\E(\phi)$ is true
exactly in the
possible world $w = v(\phi)$".\\
In order to introduce a  concept of \emph{absolute truth or falsity}
(not relative to a single possible world in $\W$) for the ground
meta formulae in $\L_F$,  we need a kind of autoepistemic modal
operator $~\diamondsuit$ (it is not part of a language $\L_F$
obtained by ontological encapsulation).  Consequently, for any given
ground formula $\Phi \in \L_F$, similarly to Moore's autoepistemic
operator, a formula $~\diamondsuit\Phi$ is able to capture the
2-valued notion of "$\Phi$ \emph{is a semantic reflection of a
many-valued logic program model} $I_{mv}$".
\\Notice that in this encapsulation, for example, the meta-implication $\leftarrow^A$ derived from the  many-valued implication, $\E(\phi)~ \leftarrow^A~\E(\psi) = \E(\phi~ \leftarrow~~\psi)$,
specifies how, for a given clause in $PR$, a logic value of the body
"propagates" to the head of this clause. It is not functionally
dependent on  the truth values of its arguments, thus it must be a
binary \emph{modal} operator. A Kripke semantics for this
 \emph{binary} modal operators can be defined based on the simple
idea  of transforming the many-valued lattice-based operator
$\rightarrow$ into the ternary accessibility relations
$\R_{\rightarrow}$.
The idea to use ternary relations to model binary modal operators
comes from  Relevance logic \cite{Urqu72,RoMe73,Dose92,Dunn93}, but,
as far as we know, this is the first time that ternary relations
have been built directly from the truth-tables for multi-valued
binary logic operators.
\begin{definition} \label{Def:K-infinite}
Let $PR$ be a many-valued logic program with a set of predicate
symbols $P$,  a many-valued Herbrand model $I_{mv}:H\rightarrow \W$
and its semantic-reflection  $\K $.\\ Then, the model of the
flattened program $PR_F$ in Definition \ref{def:flatt} is defined as
the Kripke-style model
$\M = (\W, \{\R_{\sim},\R_{\wedge}, \R_{\vee},\R_{\rightarrow}, \R_{\times} = \W \times \W \}, S, V)$, where,\\
 $\R_{\wedge} = \{(x \wedge y, x,y)~|~ x,y \in \W \}$,
 $~~\R_{\vee} = \{(x \vee y, x,y)~|~ x,y \in \W \}$,\\
 $~~\R_{\rightarrow} = \{(x \rightarrow y, x,y)~|~ x,y \in \W$ and $x \leq y
 \}$, $~~\R_{\sim} = \{(\sim x, x)~|~ x \in \W \}$,\\
  and $~~~V:\W\times P_F \rightarrow {\bigcup}_{n \in \N}
\textbf{2}^{S^n \times \W}~$ (from Definition \ref{def:KripSem}),
such that for any $p \in P$ with arity $n$ (i.e., $p_F \in P_F$ with
arity $n+1$),
 a tuple of constants $ (c_1,..,c_n) \in S^n $, and a world $w \in \W$,
$~~V(w, p_F)(c_1,..,c_n, \alpha) = 1~~$ iff $~~w = \alpha = \kappa_p(c_1,..,c_n))$,\\
  such that, for any  formula $\Phi, \Psi \in \L_F$, the satisfaction relation
$\models_{w,g}$, for a given assignment $g$ and a world $w \in \W$, is defined as follows:\\
1. $~\M \models_{w,g} p_F(x_1,..x_n,\alpha)~~$ iff
$~~V(w,p_F)(g(x_1),..,g(x_n),\alpha) = 1$.\\
2. $~\M \models_{w,g} \sim^A\Phi ~~$ iff $~~\exists y((w,y) \in
\R_{\sim}$ and $\M\models_{y,g} \Phi)$. \\
3. $~\M \models_{w,g} \wedge^A(\Phi,\Psi) ~~$ iff $~~\exists
y,z((w,y,z) \in \R_{\wedge}$ and $\M\models_{z,g} \Phi$ and
$\M\models_{y,g} \Psi)$. \\
4. $~\M \models_{w,g} \vee^A(\Phi,\Psi) ~~$ iff $~~\exists
y,z((w,y,z) \in \R_{\vee}$ and $\M\models_{z,g} \Phi$ and
$\M\models_{y,g} \Psi)$. \\
5. $~\M \models_{w,g} ~\leftarrow^A(\Phi,\Psi) ~~$ iff $~~\exists
y,z((w,y,z) \in \R_{\rightarrow}$ and $\M\models_{z,g}
\Phi$ and $\M\models_{y,g} \Psi)$. \\
6. $~\M \models_{w,g} ~\diamondsuit \Phi~~$ iff $~~\exists y((w,y)
\in \R_{\times}$ and $\M\models_{y,g} \Phi)$.
\end{definition}
The binary operators $\wedge^A, \vee^A$ and $\leftarrow^A$ for this
multi-modal logic are the existential modal operators w.r.t. the
ternary relation $\R_{\wedge}, ~\R_{\vee}$ and $ \R_{\rightarrow}$
respectively, while  $\sim^A$ and $\diamondsuit$  are the
existential unary modal operator w.r.t the binary relation
$\R_{\sim}$ and $~\R_{\times}$, respectively.\\
 Instead of
$\wedge^A(\E(\phi),\E(\psi))$, $\vee^A(\E(\phi),\E(\psi))$ and
$\leftarrow^A(\E(\phi),\E(\psi))$,   we will use also
$\E(\phi)\wedge^A \E(\psi)$, $\E(\phi)\vee^A \E(\psi)$ and
$\E(\phi)\leftarrow^A \E(\psi)$ respectively.
\begin{propo} \label{Prop:2-valued-infinite} For any assignment $g$ and a   formula $\Phi \in \L_F$ we have
that $|\diamondsuit \Phi/g| \in \{\emptyset, \W\}$, where
$\emptyset$ is the empty set. That is, for any many-valued formula
$\phi \in \L$ the formula $\diamondsuit \E(\phi/g)$ is true in the
Kripke-style relational model $\M$ given by Definition
\ref{Def:K-infinite}, so that $\M$ is a Kripke-style model of $PR_F$
correspondent to the many-valued algebraic model $I_{mv}$ of the
original program $PR$.
\end{propo}
 \textbf{Proof:} In what follows we denote by $v$ the (homomorphic) extension of a Herbrand model $I_{mv}$ to all
 ground formulae in $I_{mv}$, as defined in Definition \ref{def:valuation}.\\
  Let us demonstrate that for
any $\phi \in \L$, i.e., $\E(\phi) \in \L_F$, holds that\\
$~~~\M\models_{w,g} \E(\phi)~~$ iff $~~w = v(\phi/g)$. \\
1. For any atomic formula $p(x_1,..,x_n)$ we have that,
\\ $\M\models_{w,g} \E(p(x_1,..,x_n))~~$ iff
$~~V(w,p_F)(g(x_1),..,g(x_n),\kappa_p(g(x_1),..,g(x_n))) =1 ~~$ iff
$~~w = \kappa_p(g(x_1),..,g(x_n)) = \lambda \I(p)(g(x_1),..,g(x_n))
= I_{mv}(p(g(x_1),..,g(x_n))) = v(p(x_1,..,x_n)/g)$. Viceversa, if
 $w  = v(p(x_1,..,x_n)/g)$, i.e., $w =
I_{mv}(p(x_1,..,x_n)/g) = \kappa_p(g(x_1),..,g(x_n))$, then
$~~V(w,p_F)(g(x_1),..,g(x_n),\kappa_p(g(x_1),..,g(x_n))) =1 $ and,
consequently, from point 1 of definition above,  $~~\M\models_{w,g}
\E(p(x_1,..,x_n))$.\\
 Suppose, by the inductive hypothesis,  that $~\M\models_{z,g}
\E(\phi)~~$  iff $~~z = v(\phi/g)$, and \\$~\M\models_{y,g}
\E(\psi)~~$ iff $~~y = v(\psi/g)$, then:\\
2. For any formula $\varphi = \sim \phi$,  we have that
$~~\M\models_{w,g} \E(\varphi)~~$ iff $~~\M\models_{w,g} \E(\sim
\phi)~~$ iff $~~\M\models_{w,g} ~\sim^A\E(\phi)~~$ iff $~~(\exists
z((w,z) \in \R_{\sim}$ and $\M\models_{z,g} \E(\phi)))$, that is,
if\\ $ ~w =  ~\sim z ~~~~$ (from the definition of accessibility
relation
$\R_{\sim}$)\\ $ = ~\sim v(\phi/g) = v(\sim \phi/g)~~~~$ (from a homomorphic property of  $v$)\\
$   = v(\varphi/g)$.\\
  Viceversa, if  $~w = v(\varphi/g) = v(\sim \phi)/g =
~\sim v(\phi/g) = ~\sim z $ then, from the inductive hypothesis,
$~\M\models_{w,g} ~\sim^A \E(\phi)$, i.e.,
$~\M\models_{w,g} \E(\varphi)$. \\
3. For any formula $\varphi = \phi \odot \psi$, where $\odot \in
\{\wedge, \vee, \rightarrow\}$, we have that $~~\M\models_{w,g}
\E(\varphi)~~$ iff $~~\M\models_{w,g} \E(\phi \odot \psi)~~$ iff
($~~\M\models_{w,g} \E(\phi) \odot^A \E(\psi)~~$ iff $~~(\exists
y,z((w,y,z) \in \R_{\odot}$ and $\M\models_{z,g} \E(\phi)$ and
$\M\models_{y,g} \E(\psi)))$, that is, if\\ $ ~w =  z \odot y~~~~$
(from a definition of accessibility relation $\R_{\odot}$)\\ $ =
v(\phi/g) \odot v(\psi/g) = v(\phi/g
\odot \psi/g)~~~~$ (from a homomorphic property of  $v$)\\
$ = v((\phi \odot \psi)/g)  = v(\varphi/g)$.\\
  Viceversa, if  $~w = v(\varphi/g) = v(\phi \odot \psi)/g =
v(\phi/g) \odot v(\psi/g) = z \odot y$ then, from the inductive
hypothesis, $~\M\models_{w,g} \E(\phi) \wedge^A \E(\psi)$, i.e.,
$~\M\models_{w,g} \E(\varphi)$. \\
Thus, for any $\Phi \in \L_F$ we have that $| \Phi/g| = \{ w \}$ for
some $w\in \W$, if $\Phi/g = \E(\phi/g)$; otherwise
$| \Phi/g| = \emptyset$.\\
Consequently, we have that  $| \diamondsuit \Phi/g| = \{ w~|~
~\exists y((w,y) \in \R_{\times}$ and $\M\models_{y,g} \Phi) \} =
\W$ if $\Phi/g = \E(\phi/g)$; otherwise $| \diamondsuit \Phi/g| =
\emptyset$. That is, each ground modal formula $\diamondsuit \Phi/g$
for any $\Phi \in \L_F$ is a \emph{2-valued formula}.\\
From Definition \ref{Def:K-infinite} we have seen how a many-valued
model $I_{mv}$ of a logic program $PR$ uniquely determines a Kripke
model  $\M$ of its meta-logic modal program $PR_F$. Let us now show
the opposite direction, that is, how a Kripke model $\M$ of a modal
logic program $PR_F$ obtained by ontological encapsulation of the
original many-valued logic program $PR$, determines uniquely a
many-valued model $I_{mv}$ of the logic program $PR$. That is, let
us show that the set of ground atomic modal formulae $\diamondsuit
p_F(c_1,...,c_n, \alpha)$ for $p_F(c_1,...,c_n, \alpha) \in H_F$,
which are \emph{true} in a Kripke model $\M$, uniquely determines
 the many-valued Herbrand model
$I_{mv}$ of the original logic program $PR$:\\
In fact, we define uniquely the mapping $I_{mv}:H \rightarrow \W$,
as follows: for any modal atomic formula $\diamondsuit
p_F(c_1,...,c_n, \alpha)$, \emph{true} in the Kripke model $\M$, we
define $I_{mv}(p(c_1,....c_n))\\ = \alpha$. It is easy to verify
that such a definition of a mapping $I_{mv}:H \rightarrow \W$ is a
Herbrand model of a many-valued logic program $PR$.\\ $\square$\\
This transformation of  multi-valued logic programs into 2-valued
multi-modal logic programs can be briefly explained as follows: we
transform the original multi-valued atoms into the meta 2-valued
atoms by enlarging the original atoms with a new logic attribute
with the domain of values in $\W$. This ontological encapsulation
also eliminates  the negation (in this case the negation $\sim $) by
introducing a unary modal operator $\sim^A$. The remained binary
multi-valued lattice operations are substituted by the 2-valued
binary modal operators, by transforming the truth functional tables
of these operators directly into the ternary
accessibility relations of this modal logic.\\
\textbf{Remark:} In addition, this ontological encapsulation of
logic programs into the \emph{positive} (without the negation) modal
programs, can be used, with some opportune modifications of the
definitions above where a ground atom $p_F(c_1,..c_n,\alpha) \in
H_F$ is true only for exactly one value $\alpha \in \W$, to deal
 with the \emph{inconsistency} of 2-valued logic programs:
the resulting positive modal program will be a \emph{paraconsistent}
logic program, that is for any given ground atom $p(c_1,..c_n)$ of
the original 2-valued logic program that is inconsistent (both true
and false), in the transformed \emph{consistent} positive modal
program we can (consistently)  have two true ground atoms,
$p_F(c_1,..c_n,1)$ and
$p_F(c_1,..c_n,0)$.\\
The relationship between these two program transformations, for
finite and infinite cases of many-valued programs, can be given by
the following corollary:
\begin{corollary} For any  atom $p(x_1,..,x_n)$ of a many-valued
logic program and its two 2-valued program transformations defined
previously, the following semantic connection holds
$~~~~\M\models_{w,g} \E(p(x_1,..,x_n))~~$ iff $~~"
[w]p(g(x_1),..,g(x_n))$ is true in $\M_I"$.
\end{corollary}
 Consequently, we can conclude that  many-valued logic programs can be
equivalently replaced by  positive 2-valued multi-modal logic
programs, and this reduction of  many-valued logics into modal
logics also explains  the good properties of many-valued logic
programs.\\
Moreover, we have shown that by a 2-valued reduction of many-valued
\emph{logic programs} we obtain a 2-valued
\emph{non-truth-functional} logic, and that such a logic is just a
2-valued (multi)modal logic with a non-standard autoreferential
Kripke semantics, because modal operators are generally \emph{non
monotonic}, and in a second  case we need also \emph{binary} modal operators.\\
In what follows we will generalize this 2-valued reduction presented
for only Logic Programs, to any kind of many-valued logics.
%
\section{A general abstract reduction of many-valued into 2-valued logics}
The term "abstract" used for this general many-valued reduction
means that we do not consider any further the specific reduction of
particular functional logic operators in $\Sigma$ of a many-valued
logic into correspondent modal operators, but rather a general
reduction independent of them, based on structural consequence
operations or
matrices.\\
As we will see, both abstract reductions will result in a kind of
2-valued modal logic that are not truth-functional, as we obtained
in the specific case for Logic Programs in Section
\ref{Section:LogProg}.\\
In \cite{Susz77}  Suszko's thesis was presented. This paper is
extremely dense and very short, and thus it is not easy to
understand;  it is a kind of synthesis, in four pages, of some deep
reflections carried out by Suszko over forty years. Only 15 years
after this publication, Malinowski's book \cite{Mali93} has thrown
some light on it (see especially Chapter 10, Section 10.1).
Unfortunately, neither the quoted  paper by Suszko nor Malinowski's
book   explicitly state Suszko's thesis, but in another paper
\cite{Mali94} Malinowski has written "Suszko's thesis ... states
that each logic, i.e., a \emph{structural consequence operation}
conforming Tarski's conditions, is logically two-valued", and (p.73)
"each (structural) \emph{propositional logic} (L,C) can be
determined by a class of logical valuations of the
language $\L$ or, in other words, it is logically \emph{two-valued}".\\
In what follows we will try to formally develop  a reduction of a
many-valued \emph{predicate} logic $\L_{mv}$, with a Herbrand base
$H$, into a 2-valued logic,
based on these  observations of Suszko.\\
We denote, for a given set of thesis (ground formulae) $\Gamma$ of a
many-valued logic $\L_{mv}$, the \emph{2-valued} structural
consequence relation by $\Gamma \vdash \phi$, which means that a
ground formula $\phi$ is a structural consequence of set of ground
formulae in $\Gamma$, i.e., that $\phi \in C(\Gamma)$ where $C$ is a
structural consequence operation conforming Tarski's conditions. \\
We denote by $Val = \B^H$ the set of Herbrand many-valued
interpretations $v:H \rightarrow \B$, $v \in Val$, for a many-valued
logic $\L_{mv}$ with a Herbrand base $H$ and a set of
\emph{algebraic} truth-values in $\B$. Let $Val_{\Gamma} \subset
Val$ be a non-empty subset of \emph{models} of $\Gamma$, that is,
valuations $v \in Val_{\Gamma}$
that satisfy every ground formula in $\Gamma$.\\
Then, the truth of $\Gamma \vdash \phi$  is equivalent to the fact
that  every valuation $v \in Val_{\Gamma}$ is a model of $\phi$ also
(i.e., satisfies a ground formula $\phi$). However, here we are not
speaking about a truth value of a many-valued ground formulae $\phi
\in \L_{mv}$, but about a truth value of a meta sentence $\Gamma
\vdash \phi$. In what follows, for a fixed set of (initial) thesis
$\Gamma \subset \L_{mv}$ that  defines a structural many-valued
logic $(\Gamma, C)$, we will transform the left side construct
$\Gamma \vdash (\_)$ in an universal \emph{modal operator}
$\Box_{\Gamma}$ ("$\Gamma$-deducible"), so that a meta sentence
$\Gamma \vdash \phi$ can be replaced by an equivalent modal
formula $\Box_{\Gamma} \phi$ in this 2-valued meta logic.\\
Thus, analogously to the more specific cases for Logic Programs,
also in this general abstract 2-valued reduction we are not speaking
about the two-valuedness of an original \emph{many-valued formula},
but about a \emph{modal formula} of a 2-valued meta-logic obtained
by this
transformation.\\
What remains now is to define a Kripke semantics for this modal
meta-logic, denoted by $\L_{\F}$, obtained from a set of formulae
$\F = \{\Box_{\Gamma} \phi ~|~ \phi \in \L_{mv} \}$ and the standard
2-valued logic connectives (conjunction, disjunction, implication
and negation).
\begin{definition} \label{Def:Suszko} Given a structural many-valued logic $(\Gamma, C)$, where $\Gamma
\subset\L_{mv}$ is a subset of ground formulae with a set of
predicate symbols in $P$ and a Herbrand base $H$, we define  a
 Kripke-style model for Suszko's reduction, $~\M = (\W, \R_{\Gamma}, S, V)$, where a
set of possible worlds is $\W = Val$, $\R_{\Gamma} = Val \times
Val_{\Gamma}$,
   and $~~~V:\W\times P \rightarrow {\bigcup}_{n \in \N}
\textbf{2}^{S^n \times \W}~$ (from Definition \ref{def:KripSem}),
such that for any $p \in P$ with arity $n$,
 a tuple of constants $ (c_1,..,c_n) \in S^n $, and a world $w \in
 \W$, (a Herbrand interpretation $w:H \rightarrow \B$),
$~~V(w, p)(c_1,..,c_n) = 1~~$ iff $~~w \in Val_{\Gamma}$.\\
  The satisfaction relation
$\models_{w,g}$, for a given assignment $g$ and a world $w \in \W$, for any  many-valued formula $\phi, \psi $,  is defined as follows:\\
1. $~\M \models_{w,g} p(x_1,..x_n)~~$ iff
$~~V(w,p)(g(x_1),..,g(x_n)) = 1$.\\
2. $~\M \models_{w,g} \phi~~$ iff
$~~$ the homomorphic extension (in Definition \ref{def:valuation}) of the Herbrand model $w$ is a model of the ground formula $\phi/g$.\\
3. $~~{\M} \models_{w,g}~\Box_{\Gamma} \phi~~~$ iff $~~~\forall
w'((w,w') \in {\R}_{\Gamma} $ implies ${\M} \models_{w',g}~
\phi~)~$.\\
4. $~~{\M} \models_{w,g}~ \neg \Box_{\Gamma}\phi ~~~$ iff  $~~~$ not
$~{\M} \models_{w,g}~
\Box_{\Gamma} \phi~$ ,\\
 5. $~~{\M} \models_{w,g}~ \Box_{\Gamma}\phi \wedge \Box_{\Gamma}\psi~~~$ iff $~~~{\M}
\models_{w,g}~ \Box_{\Gamma} \phi~$ and $~{\M} \models_{w,g}~
 \Box_{\Gamma} \psi~$,\\
 6. $~~{\M} \models_{w,g}~ \Box_{\Gamma}\phi \vee \Box_{\Gamma}\psi~~~$ iff $~~~{\M} \models_{w,g}~ \Box_{\Gamma}\phi~$ or $~{\M} \models_{w,g}~
 \Box_{\Gamma}\psi~$,\\
 7. $~~{\M} \models_{w,g}~ \Box_{\Gamma}\phi \rightarrow \Box_{\Gamma}\psi~~~$ iff $~~~{\M} \models_{w,g}~ \Box_{\Gamma}\phi~$ implies $~{\M} \models_{w,g}~
 \Box_{\Gamma}\psi~$,\\
where the logic connectives $\wedge, \vee, \rightarrow$ and $\neg$
are the classic 2-valued conjunction, disjunction, implication and
negation respectively.
\end{definition}
Notice that a satisfaction of the 2-valued formulae of this
meta-logic $\L_{\F}$, obtained by Suszko's reduction of the original
many-valued logic, is relative to points 3 to 7 in the Definition
above. Consequently, the two-valuedness is a property not of the
original many-valued formulae, but of the modal formulae in this non
truth-functional modal meta-logic. Let us show that this reduction
is  sound and complete.
\begin{lemma} \label{Lemma:Suszko} Given a Kripke model $~\M = (\W, \R_{\Gamma}, S, V)$
in Definition \ref{Def:Suszko}, for a given many-valued logic
$(\Gamma, C)$, where $\Gamma \subset\L_{mv}$ is a subset of ground
formulae, then for any formula $\phi \in \L_{mv}$ and assignment $g$
we have that:\\ $~~~\phi/g \in C(\Gamma)$, (i.e., $\Gamma \vdash
\phi/g)~~~$ iff $~~~ \Box_{\Gamma} \phi/g$ is true in $\M$.
\end{lemma}
\textbf{Proof:} If $\Gamma \vdash \phi/g~$ then for every $w \in
Val_{\Gamma}$ its homomorphic extension to all ground formulae in
$\L_{mv}$ is a model of a ground formula $\phi/g \in \L_{mv}$.
Thus,\\ $|\Box_{\Gamma}\phi/g| = \{w~|~ \forall w'((w,w') \in
{\R}_{\Gamma} $ implies ${\M} \models_{w',g}~ \phi~)~\} \\= \{w~|~
\forall w'((w,w') \in {\R}_{\Gamma} $ implies $w'$ is a model of
$\phi/g~)~\} \\= \{w~|~ \forall w'(w' \in Val_{\Gamma} $ implies
$w'$ is a model of $\phi/g~)~\} \\ = \{w~|~$  true $~\} = \W$,
$~~~~$
i.e., $\Box_{\Gamma}\phi/g$ is true in $\M$.\\
Viceversa, if $\Box_{\Gamma}\phi/g$ is true in $\M$ then $\W =
|\Box_{\Gamma}\phi/g| = \{w~|~ \forall w'((w,w') \in {\R}_{\Gamma} $
implies ${\M} \models_{w',g}~ \phi~)~\} = \{w~|~ \forall w' \in
Val_{\Gamma} $ ($w'$ is a model of $\phi/g~)~\}$, that is, the
following sentence has to be true: $~ \forall w' \in Val_{\Gamma} $
($w'$ is a model of $\phi/g~)$, and, consequently, $\Gamma \vdash
\phi/g~$, i.e., $~\phi/g \in C(\Gamma)$.
\\ $\square$\\
These results confirm da Costa's idea \cite{KoCo80} that a reduction
to 2-valuedness can be done at an abstract level, without taking
into account the underlying structure of the set of many-valued
formulae (differently from the particular case of Logic Programs
given in Section \ref{Section:LogProg}). \\It is not necessary to
make a detour by matrices in order to get this reduction. But in the
case where we have a many-valued logic with a given matrix $(\B,
D)$, where $D \subset \B$ is a subset of designated algebraic truth
values, then we are able to define a new modal 2-valued reduction
for such a many-valued logic, based on the \emph{existential} modal
operator $\lozenge_D$ ("$D$-satisfied"). It is given in the way
that, for given homomorphic extension of a valuation $v:H
\rightarrow \B$, a many-valued formula $\phi \in \L_{mv}$ and an
assignment $g$, the formula $\lozenge_D \phi/g$ is true $~~$
iff $~~v(\phi/g) \in D$, that is,$~~$ iff $~~v$ satisfies (is \emph{a model} of)  $\phi/g$.\\
What remains now is to define a Kripke semantics for this
matrix-based reduction to a modal meta-logic, denoted by $\L_{\E}$,
obtained from a set of formulae $\E = \{\lozenge_{D} \phi ~|~ \phi
\in \L_{mv} \}$ and standard 2-valued logic connectives
(conjunction, disjunction, implication and negation).
\begin{definition} \label{Def:matrix} Given a  many-valued logic $\L_{mv}$ with a given matrix $(\B, D)$,
 a set of predicate symbols in $P$ and a Herbrand base $H$,
we define  a  Kripke-style model for a matrix-based reduction by a
quadruple $~\M = (\W, \R_{D}, S, V)$, where a set of possible worlds
is $\W = \B$, $\R_{D} = \B \times D$,
   and $~~~V:\W\times P \rightarrow {\bigcup}_{n \in \N}
\textbf{2}^{S^n \times \W}~$ (from Definition \ref{def:KripSem}),
such that for any $p \in P$ with arity $n$,
 a tuple of constants $ (c_1,..,c_n) \in S^n $, \\
$~~V(w, p)(c_1,..,c_n) = 1~~$ for exactly one world $w \in D
\subseteq \W$.\\
  The satisfaction relation
$\models_{w,g}$, for a given assignment $g$ and a world $w \in \W$, for any  many-valued formula $\phi, \psi \in \L_{mv} $,  is defined as follows:\\
1. $~\M \models_{w,g} p(x_1,..x_n)~~$ iff
$~~V(w,p)(g(x_1),..,g(x_n)) = 1$.\\
2. $~\M \models_{w,g} \phi~~$ iff $~~w = v(\phi/g) \in D$, where $v$
is the unique homomorphic extension (Definition \ref{def:valuation})
of a mapping $v:H \rightarrow \B$ defined by: for each
$p(c_1,...,c_n) \in H$, $v(p(c_1,...,c_n)) = y$ such that $~V(y,
p)(c_1,..,c_n) = 1$.
\\
3. $~~{\M} \models_{w,g}~\lozenge_{D} \phi~~~$ iff $~~~\exists
w'((w,w') \in {\R}_{D} $ and ${\M} \models_{w',g}~
\phi~)~$.\\
4. $~~{\M} \models_{w,g}~ \neg \lozenge_{D}\phi ~~~$ iff  $~~~$ not
$~{\M} \models_{w,g}~ \lozenge_{D} \phi~$ ,\\
 5. $~~{\M} \models_{w,g}~ \lozenge_{D}\phi \wedge \lozenge_{D}\psi~~~$ iff $~~~{\M}
\models_{w,g}~ \lozenge_{D} \phi~$ and $~{\M} \models_{w,g}~ \lozenge_{D} \psi~$,\\
 6. $~~{\M} \models_{w,g}~ \lozenge_{D}\phi \vee \lozenge_{D}\psi~~~$ iff $~~~{\M} \models_{w,g}~ \lozenge_{D}\phi~$ or $~{\M} \models_{w,g}~
 \lozenge_{D}\psi~$,\\
 7. $~~{\M} \models_{w,g}~ \lozenge_{D}\phi \rightarrow \lozenge_{D}\psi~~~$ iff $~~~{\M} \models_{w,g}~ \lozenge_{D}\phi~$ implies $~{\M} \models_{w,g}~
 \lozenge_{D}\psi~$,\\
where the logic connectives $\wedge, \vee, \rightarrow$ and $\neg$
are the classic 2-valued conjunction, disjunction, implication and
negation respectively.
\end{definition}
Notice that in this case we obtained an autoreferential semantics
\cite{Majk06ml,MaPr09} and that a satisfaction of the 2-valued
formulae of this meta-logic $\L_{\E}$, obtained by the matrix-based
reduction of original many-valued logic, is relative to points  3 to
7 in the Definition above. Consequently, the two-valuedness is a
property not of the original many-valued formula, but of the modal
formula in this non truth-functional modal meta-logic.\\ Let us show
that this matrix-based reduction is  sound and complete.
\begin{lemma} \label{Lemma:matrix} Let $~\M = (\W, \R_{D}, S,
V)$ be a Kripke model, given in Definition \ref{Def:matrix}, for a
many-valued logic $\L_{mv}$ with a matrix $(\B,D)$. We define a
many-valued Herbrand interpretation $v:H \rightarrow \B$ as follows:
for each $p(c_1,...,c_n) \in H$,\\ $v(p(c_1,...,c_n)) = w$, where
$w$ is the unique value that
satisfies  $V(w,p)(c_1,...,c_n) = 1$.\\
 Then, for any
formula $\phi \in \L_{mv}$ and an assignment $g$,  we have that, \\
$~~~$" the homomorphic extension of $~v~$ is a model of $~\phi/g "
~~~$ iff $~~~ \lozenge_{D}\phi/g$ is true in $\M$.
\end{lemma}
\textbf{Proof:} If the homomorphic extension of $~v~$ is a model of
$~\phi/g$ then  $w' = v(\phi/g) \in D$, thus,
$~~~|\lozenge_{D}\phi/g| = \{w~|~ \exists w'((w,w') \in {\R}_{D} $
and ${\M} \models_{w',g}~ \phi~)~\} \\= \{w~|~ \exists w'((w,w') \in
{\R}_{D} $ and $w'= v(\phi/g)~\}  = \{w~|~$  true $~\} = \W$,\\
i.e., $\lozenge_{D}\phi/g$ is true in $\M$.\\
Viceversa, if $\lozenge_{D}\phi/g$ is true in $\M$ then $\W =
|\lozenge_{D}\phi/g| = \{w~|~ \exists w'((w,w') \in {\R}_{D} $ and
${\M} \models_{w',g}~ \phi~)~\} = \{w~|~ \exists w' \in D $ ($w'=
v(\phi/g))~\}$, that is, the following sentence has to be true: $~
\exists w' \in D $ ($w'= v(\phi/g))$, and, consequently, it must
hold that $ v(\phi/g) \in D$, i.e., the homomorphic extension of
$~v~$ is a model of $~\phi/g$.
\\ $\square$

\section{Conclusion}
As we mentioned,  real-world problems often have to be resolved by
applying  Artificial Intelligence techniques by means of many-valued
logics (fuzzy,  paraconsistent, bilattice-based, etc..), therefore,
the investigation of the general properties of  these non standard
many-valued logics is a very important issue. Based on Suszko's
thesis, in this paper we analyzed a different possibility of
reducing these many-valued logics into 2-valued logics, in order to
be able to compare their original many-valued properties based on
such obtained 2-valued logic. Our approach, however, is formal and
constructive, in contrast to Suszko's nonconstructive approach based
on a distinction between
designated and undesignated algebraic truth-values. \\
We introduced a kind of a contextualization for many-valued logics
that is similar to the special   annotated logics case, but which
gives us the possibility of continuing to use the standard Herbrand
models as well. In this paper we have shown how many-valued logic
programs can be equivalently transformed into  contextual logic
programs with higher-order Herbrand interpretations. We have shown
that the flattening of such higher-order Herbrand interpretations
leads to 2-valued logic programs, identical to meta logic programs
obtained by an ontological encapsulation of the original Many-valued
logic programs \cite{Majk04on,MajkC04}  with modal logic
connectives. From the other side, the properties of higher-order
Herbrand types, with a possibility of introducing the Kripke
semantics for them, are the basis for an equivalent transformation
of many-valued Logic Programs into the
 2-valued multi-modal Logic Programs with modal
atoms.\\
We also developed  a general abstract 2-valued reduction for any
kind of many-valued logics, based on informal Suszko's thesis, and
have shown the Kripke semantics for obtained 2-valued modal
meta-logics, for both Suszko's (non-matrix) and
matrix-based cases.\\
Consequently, any kind of reduction of a many-valued logic into
2-valued logic results in a non truth-functional modal meta-logic,
which obviously is not an original "reference" many-valued logic.
This process is explained by the fact that this reduction is based
on new sentences about the original many-valued sentences, and that,
by avoiding the second order syntax of these meta-sentences, what is
required is the introduction of new \emph{modal} operators in this
equivalent but 2-valued meta-logic. As presented in the case of
Logic Programming and general structural many-valued logics, this is
a general approach
to 2-valued reductions.\\
  This
results consolidate an intuition that the many-valued logics, used
for uncertain, approximated and context-dependent information, can
be embedded into multi-modal logics with possible world semantics,
which are well investigated sublanguages of the standard First-order
logic language with very useful properties.\\
This method can be used for \emph{paraconsistent} logics as well, as
shown in an example for the 4-valued Belnap' bilattice, and explains
why the paraconsistent logics can be formalized by modal logics as well.\\
Further investigation: It is well known, by Definition 2 in
\cite{Majk08dC,Majk11}, that any 2-valued modal logic can be
equivalently transformed into a truth-valued many-valued logic with
a complete distributive lattice of its "algebraic functional" logic
values (so called complex algebras over powerset of possible
worlds), as for example the complex algebra for a (modal)
intuitionistic logic is a Heyting algebra over the powerset of
possible worlds. Here we demonstrated that, additionally,  every
truth-functional many-valued logic can be reduced into a non truth
functional \emph{modal} (meta) logics. There does remain an open
question: are all 2-valued non truth-functional logics
\emph{necessarily modal} logics? Consider, for example, the
paraconsistent da Costa's $C_n$ system \cite{Costa74} for which
  the relational Kripke semantics has not still been defined.

\bibliographystyle{IEEEbib}
\bibliography{mydb}
%


\end{document}